\documentclass[%
 aip,pop,
 amsmath,amssymb,
 reprint,%
]{revtex4-1}

\usepackage[utf8]{inputenc}
\usepackage{textcomp}
\usepackage{amsbsy}
\usepackage{amsfonts}
\usepackage{mathrsfs}
\usepackage{amsthm}
\usepackage{xcolor}
\usepackage{textcomp}
\usepackage{siunitx}  
\DeclareSIUnit\permille{\text{\textperthousand}}
\usepackage{color}

\usepackage[english]{babel}	 

\usepackage{graphicx}
\usepackage{dcolumn}
\usepackage{bm}

\usepackage[T1]{fontenc}
\usepackage{mathptmx}

\begin{document}

\title{The Dependence of the Impurity Transport on the Dominant Turbulent Regime in ELM-y H-mode Discharges
}

\author{T.\,Odstr\v{c}il}
 \email{odstrcil@mit.edu}
\affiliation{ Massachusetts Institute of Technology, 77 Massachusetts Avenue, Cambridge, MA USA}%
\author{ N.T. Howard}
\affiliation{ Massachusetts Institute of Technology, 77 Massachusetts Avenue, Cambridge, MA USA}%
\author{ F. Sciortino}
\affiliation{ Massachusetts Institute of Technology, 77 Massachusetts Avenue, Cambridge, MA USA}%
\author{C. Chrystal}
\affiliation{ General Atomics, PO Box 85608, San Diego, CA 92186-5608, USA}%
\author{C. Holland}
\affiliation{ University of California – San Diego, 9500 Gilman Drive, La Jolla, CA, USA}%
\author{E. Hollmann}
\affiliation{ University of California – San Diego, 9500 Gilman Drive, La Jolla, CA, USA}%
\author{G. McKee }
\affiliation{University of Wisconsin – Madison, Wisconsin 53706}%
\author{K.E. Thome}
\affiliation{ General Atomics, PO Box 85608, San Diego, CA 92186-5608, USA}%
\author{ T.M. Wilks}
\affiliation{ Massachusetts Institute of Technology, 77 Massachusetts Avenue, Cambridge, MA USA}%

\date{\today}

\begin{abstract}

Laser blow-off injections of aluminum and tungsten have been performed on the DIII-D tokamak to investigate the variation of impurity transport in a set of dedicated ion and electron heating scans with a fixed value of the external torque.  
The particle transport is quantified via the Bayesian inference method, which,  constrained by a combination of a charge exchange recombination spectroscopy, soft X-ray measurements, and VUV spectroscopy provides a detailed uncertainty quantification of the transport coefficients. Contrasting discharge phases with a dominant electron and ion heating reveal a factor of 30 increase in midradius impurity diffusion and a 3-fold drop in the impurity confinement time when additional electron heating is applied.  Further, the calculated stationary aluminum density profiles reverse from peaked in electron heated to hollow in the ion heated case, following a similar trend as electron and carbon density profiles. Comparable values of a core diffusion have been observed for W and Al ions, while differences in the propagation dynamics of these impurities are attributed to pedestal and edge transport.  Modeling of the core transport with non-linear gyrokinetics code  CGYRO [J. Candy and E. Belly J. Comput. Phys. 324,73 (2016)], significantly underpredicts the magnitude of the variation in Al transport. 
The experiment demonstrates a 3-times steeper increase of impurity diffusion with additional electron heat flux and 10-times lower diffusion in ion heated case than predicted by the modeling. However, the CGYRO model correctly predicts that the Al diffusion dramatically increases below the linear threshold for the transition from the ion temperature gradient (ITG) to trapped electron mode (TEM).  
 
\end{abstract}

\maketitle 

\section{Introduction}
Controlling of impurities in the plasma is the critical issue for current tokamaks with metal walls as well as for future fusion reactors. While low-Z impurities mostly dilute the main fuel ions, the line emission of partially stripped high-Z impurities like tungsten significantly increases the radiative cooling of the plasma even at low concentration  $c_W > 10^{-5}$ as reported on ASDEX Upgrade\cite{neu2002impurity}. Moreover, when unfavorable transport conditions occur, the high-Z ions accumulating on-axis can trigger a radiative collapse of the plasma. Excessive radiative cooling thus represents a stringent condition for a maximum tolerable level of impurities in the reactor\cite{putterich2019determination}. 
High-Z ions will inevitably be present in the reactors as intrinsic impurities from the metal wall, or they will be deliberately introduced to build up a radiative mantel at the plasma edge to protect the plasma-facing components from an excessive heat load. Therefore, it is essential to understand the impurity transport and develop strategies to control impurities with the optimal use of scarce resources like heating sources and other actuators. 

One  efficient strategy for expelling high-Z impurities from the plasma is  central wave heating by electron cyclotron heating (ECH) or ion cyclotron resonance heating (ICRH). This effect is well documented in most experiments equipped with these heating sources: Alcator C-Mod\cite{rice2015core}, \mbox{DIII-D}\cite{gohil2003recent}, TCV\cite{scavino2004effects},  JET\cite{pasini1990impurity, puiatti2003simulation, nave2003role, carraro2007impurity,giroud2007study,valisa2011metal,puiatti2006analysis}, Tore Supra\cite{villegas2010experimental}, ASDEX \cite{angioni2017comparison, neu2002impurity, dux2003influence, sertoli2011local} and recently also KSTAR\cite{hong2015control}, \mbox{HL-2A}\cite{zhang2016investigation,cui2018study} and EAST\cite{zhang2017suppression,shen2019suppression}. On-axis ECH heating is reported to flatten main ion gradients\cite{dux2005impurity} and thus reduce neoclassical inward pinch for impurities as well as increase anomalous diffusion\cite{dux2003influence, sertoli2011local,hong2015control,zhang2016investigation} to compete with the neoclassical convection.   ICRH heating is generally less efficient in increasing the anomalous diffusion\cite{dux2003influence,valisa2011metal} and more heating power is required to match ECH\cite{angioni2017comparison}. However, on-axis ICRH significantly reduces  inward pinch and with  sufficient power reverses  convection in the inner core\cite{valisa2011metal,carraro2007impurity,giroud2007study}. A plausible explanation for the reversed convection is an increased neoclassical ion temperature screening combined with ion density flattening and
reduction in the magnitude of the neoclassical flux by fast ions counter-acting a centrifugal asymmetry of high-Z impurities\cite{casson2014theoretical,bilato2016impact,odstrcil2017physics}. The neoclassical origin of this reversal is also supported by its strong Z dependence\cite{giroud2007study}.   Despite a large variation in on-axis peaking, only a minor flattening of impurity profiles with non-zero ICRH power is observed on midradius\cite{valisa2011metal,carraro2007impurity,giroud2007study,puiatti2006analysis}.    An additional mechanism responsible for expelling impurities  is a saturated $n/m = 1/1$ mode destabilized by excessive ECH power inside of $q = 1$ surface\cite{leigheb2007molybdenum,gude2010hollow,sertoli2011local,sertoli2015modification,sertoli2015interplay, angioni2017comparison,cui2018study}, driving a strong outward pinch proportional to impurity charge $Z$ inside of a displaced core.  A description of this phenomena by available transport models in non-axisymmetrical geometry\cite{garcia2017electrostatic,bergmann2016effect,ferrari2019effect} is still incomplete. 
Not all experiments demonstrate a beneficial effect of ECH heating; for instance, purely ECH heated L-mode discharges in Tore Supra\cite{villegas2010experimental} and TCV\cite{scavino2004effects}. In the former,  the diffusion decreases for on-axis ECH, while in the latter, the impurity over energy confinement time increases up to 5 for a higher ECH power.

Although a significant impact of the wave heating on diffusion was widely recognized in these experiments, most of the gyrokinetics validation effort focuses on impurity density peaking for zero particles flux\cite{angioni2011gyrokinetic,  sertoli2011local,puiatti2006analysis,valisa2011metal, casson2013validation} or combination of a neoclassical and gyrokinetics modeling compared with stationary impurity density profiles\cite{angioni2016gyrokinetic,casson2014theoretical,angioni2017comparison}. A different approach proposed in a  nonlinear gyrokinetic study\cite{angioni2015gyrokinetic} sheds light on the question of whether the increased diffusion is a mere power degradation by localized heating or ion and electron heating contributes differently. Gyrokinetics simulations, supported by an analytical model, indicated a peaking of impurity diffusion for the electron heat flux slightly exceeding the ion heat flux. The experimental validation of this prediction and quantitative comparison with gyrokinetics simulations are thus the main goals of this paper.  

To meet these objectives, it is essential to maximize a variation in electron to ion heat flux ratio while maintaining optimal conditions for an assessment of anomalous impurity transport by laser blow-off (LBO) technique. Therefore, our experimental scenario is designed with the ``predict-first'' approach, combining TRANSP \mbox{PT-SOLVER}\cite{budny2012ptransp} for prediction of time-evolving kinetics profiles using the TGYRO solver\cite{candy2009tokamak} combining TGLF\cite{staebler2007theory} and NEO\cite{belli2008kinetic} for impurity transport coefficient and the STRAHL\cite{dux2014strahl} code for synthetic impurity diagnostics data. Low collisionality, necessary for decoupling of electron and ion fluids and increasing variation in ion and electron heat fluxes, is achieved by minimizing the deuterium gas puff source to keep electron density low and increasing electron temperature by a sufficient ECH or NBI (neutral beam injection) power. The elevated electron temperature is also favorable for charge exchange spectroscopy of fully-stripped Al ion and measurements by soft X-ray diagnostics. Further, the extent of a low magnetic shear region near the magnetic axis, dominated by neoclassical transport\cite{giannella1994role}, is reduced by increasing edge safety factor $q_{95}$. Additional benefits of a higher $q_{95}$ are low core MHD activity and reduced particle confinement.  Finally, a moderate and steady plasma rotation desired to suppress $E\times B$ stabilization is attained by counter-current NBI. In these conditions,  TGYRO predicts a 5-fold increase of diffusion when ECH is applied, well outside of the expected experimental and modeling uncertainty.

The rest of this paper is divided into four parts. Section~\ref{sec:setup} gives a description of the experimental scenario, impurity diagnostics, heat transport, and power balance heat fluxes.  Detailed investigation of aluminum and tungsten transport is presented in section~\ref{sec:experimentAl}. These experimental results are contrasted with nonlinear and quasilinear gyrokinetics modeling in section~\ref{sec:gyro}, and conclusions are drawn in section~\ref{sec:end}.
Finally, in the appendix is described our forward model for impurity transport modeling, and the new Bayesian method for inference of the impurity transport coefficient.

\section{Experimental setup}
\label{sec:setup}
\begin{figure}[h]
 \centering
 \includegraphics[scale=.80]{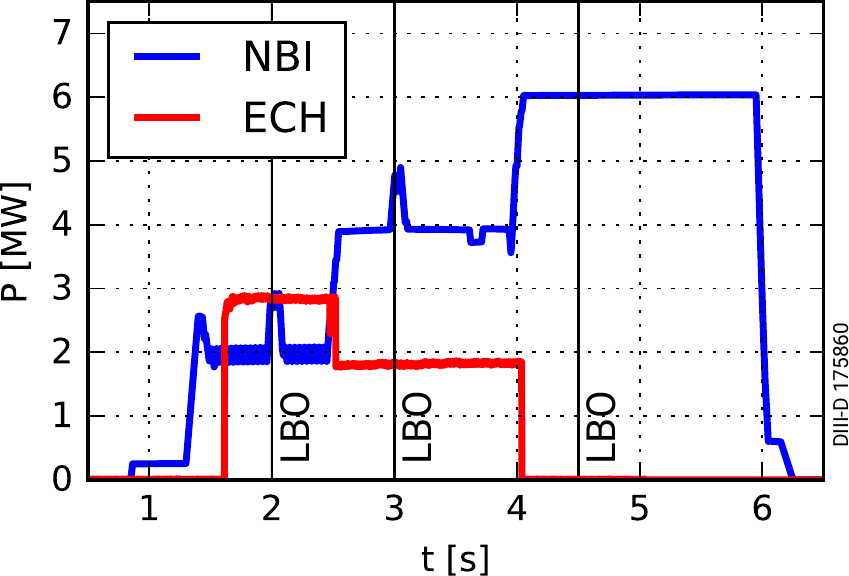}
 \caption{Timetraces of NBI (blue) and ECH (red) powers  for discharge \#175860 with times of impurity ablations indicated by  full vertical lines.  }
 \label{fig:heating_175860}
\end{figure}

\vspace{-.2cm}
 We have performed a series of heating power scans in   lower single null type-I ELMy H-mode discharges in the DIII-D tokamak\cite{luxon2002design} with the following parameters: major radius $R_0=1.77$\,m, minor radius $a=0.57$\,m, plasma current $I_p=0.9$\,MA, toroidal magnetic field on-axis $B_T=-1.97$, and safety factor \mbox{$q_{95}=5.7$}. A moderate value of $q_{95}$ allowed for sawtooth-free operation and reduced particle confinement. Further, the deuterium gas fueling was switched off immediately after the L-H transition to lower the electron density.  In the flattop of the discharge, the value of line averaged density $\bar{n}_e = 3.8\cdot 10^{19}$\,m$^{-3}$ (0.5 of the Greenwald density) was constant because of neutrals recycling from the walls and gradually increasing fueling from  NBI during the power scan.  The plasma heating was provided by a combination of NBI and near axis ECH. The ECH resonance was located on the high field side (HFS) of the plasma at $\rho_\mathrm{tor} = 0.25$, where $\rho_\mathrm{tor}$ stands for a normalized toroidal flux coordinate. The heating position was deliberately placed just outside of the $q=1$ surface to avoid destabilizing the saturated 1/1 mode, strongly reducing impurity density in the innermost part of the plasma.  The NBI heating scan started at 1.2\,s with 2\,MW of power  (see Fig.~\ref{fig:heating_175860}), followed by 2.9\,MW of ECH at 1.6\,s immediately triggering the L-H transition. At 2.5\,s, the NBI power was increased to 4.0\,MW, while the ECH power was reduced to 1.8\,MW, and in the last phase from 4.0\,s was applied 6\,MW of NBI. 
 
 The variation in the plasma rotation during the NBI power scan, which would enhance both $E\!\times\! B$ shear and poloidal asymmetries of impurities were eliminated by applying counter-current NBI to fix the input torque at 2.0\,Nm. Large and irregular edge localized modes (ELMs) were present in all three heating phases, and the mean ELM frequency doubled from 80\,Hz in the first phase up to 150\,Hz in the last phase. Since the sawteeth were absent, the only observed core MHD activities were fast ion driven fishbones and benign 3/2 neoclassical tearing mode (NTM) destabilized in the last heating phase by an increase of $\beta_N$. The NTM mode width $W$ estimated from fluctuation amplitude  $\delta T_e$ of electron cyclotron emission (ECE) diagnostic is $\delta T_e/|\nabla T_e| \sim 1$\,cm and, except for a small drop in the $T_e$ gradient at the mode location $\rho_\mathrm{tor} = 0.3$, no effect on other kinetics profiles or the impurity transport was observed. 

\begin{figure*}
 \centering
 \includegraphics[scale=.65, trim={0 0.5cm 0 0},clip]{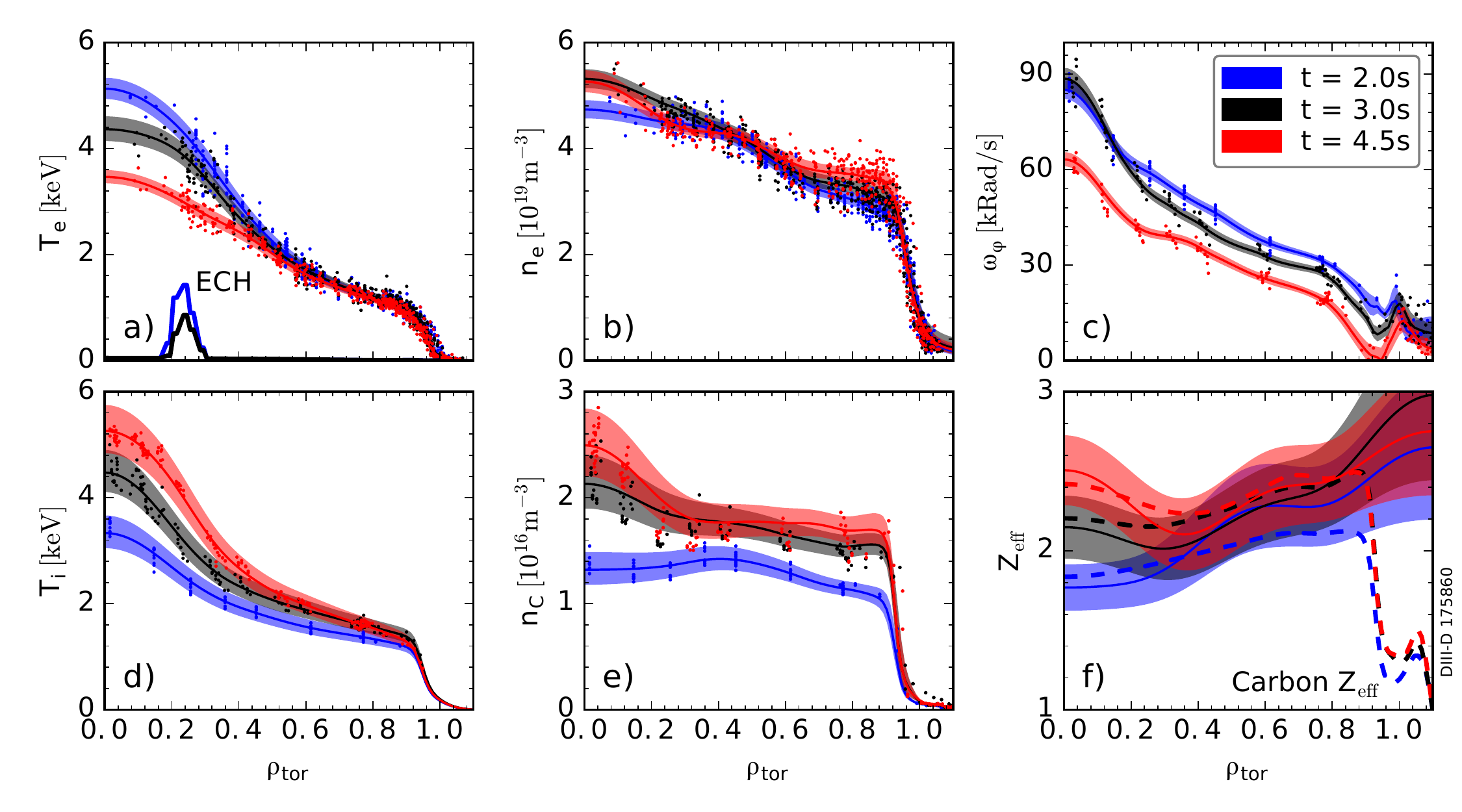}
 \caption{ Electron temperature a) and density b) from a Thomson scattering (TS), toroidal rotation c) and ion temperature d) and carbon density from a charge exchange recombination (CER), $Z_\mathrm{eff}$ from a visible bremsstrahlung f) and  density of carbon $C^{6+}$ (dashed line). Fits are averaged over 50\,ms after each impurity injected at 2.0, 3.0, and 4.5\,s, and dots correspond to the actual measurement. Peaks at the bottom of a) represent ECH deposition profiles.}
 \label{fig:175860_kinprof}
\end{figure*}

\subsection{Kinetics profiles}
\begin{figure*}
 \centering
 \includegraphics[scale=.65]{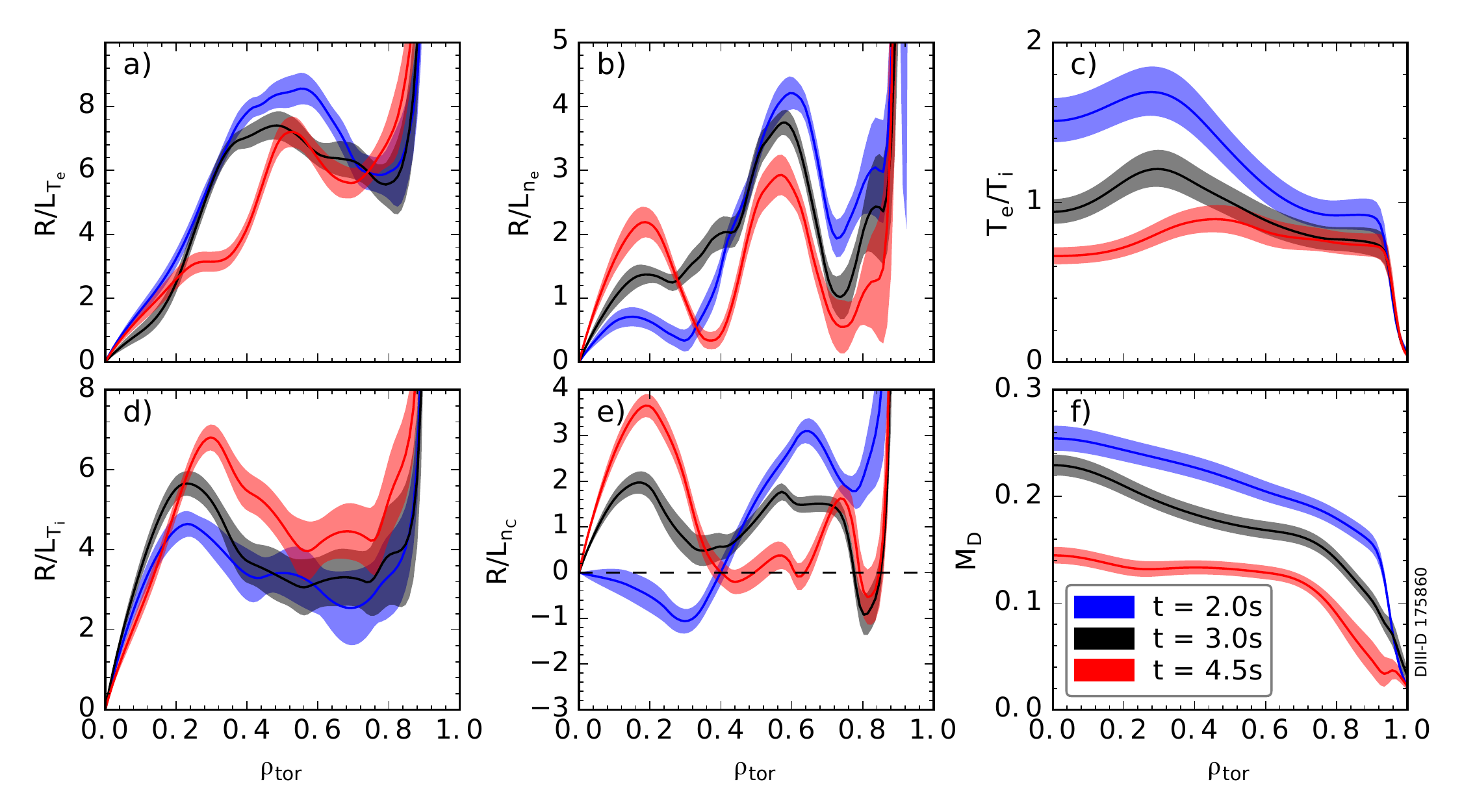}
 \caption{ Normalized gradient lengths of $T_e$ a), $n_e$ b), $T_i$ d) and $n_C$ e) corresponding to profiles in Fig.~\ref{fig:heating_175860}. The $T_e/T_i$ ratio and deuterium Mach number $M_D$ are plotted in c) and f) respectively.   }
 \label{fig:175860_derived}
\end{figure*}

\vspace{-.3cm}
Fits of the radial profiles, including uncertainties presented in Fig.~\ref{fig:175860_kinprof}, as well as their gradients in Fig.~\ref{fig:175860_derived}  are inferred via 2D Gaussian process regression (GPR)  of experimental data.  Detailed investigation of the profile variation during the three heating phases reveals the impact of the heating scheme on the plasma's kinetic profiles. A significant variation in the $T_e$ and $T_i$ profiles is a direct consequence of applied central electron heating and its relative balance with the ion heating. Electron temperature profiles $T_e$ measured by Thomson scattering (TS)\cite{ponce2010thomson} sharply peak inside of $\rho_\mathrm{tor} < 0.5$ during the ECH phase while outside this region, the $T_e$ profiles are unchanged through the heating scan. The ion temperature profile $T_i$ from charge exchange recombination (CER)\cite{chrystal2016improved} is moderately peaked in the ECH phase, and the peaking increases with more delivered NBI  power. 
The radial profiles of carbon toroidal angular velocity $\omega_\varphi$ are generally the same for two ECH cases whereas the NBI-only case has systematically lower $\omega_\varphi$.  Considering the same external momentum input is present in all cases, this indicates a change in momentum transport or a different level of intrinsic torque at the plasma edge. The electron density $n_e$ in the ECH phases is flat near of the magnetic axis, but the density gradient $R/L_{n_e}$ peaks in midradius in Fig.~\ref{fig:175860_derived}b, in opposite to the commonly observed density ``pump-out'' . This is in contrast to the NBI phase, where the density gradient is larger near axis and smaller on midradius. Despite the variation in gradients, the midradius density is the same in all cases due to the growth of a pedestal top  $n_e$ with NBI power. The same core peaking is also present in a carbon density $n_C$ and an effective charge $Z_\mathrm{eff}$ inferred from line-integrated visible bremsstrahlung \cite{schissel1988measurements,callahan2019integrated}. $Z_\mathrm{eff}^C$~estimated from $n_C$ matches $Z_\mathrm{eff}$ from the visible bremsstrahlung, indicating that the carbon is a dominant light impurity in the plasma. The same conclusion is drawn from a soft X-ray (SXR) data (not shown) where  D+C bremsstrahlung contributed to $\sim$90\% of the measured SXR emissivity before impurity injection except for the NBI heated case, where a significant peaking of intrinsic high-Z impurities inside of $\rho_\mathrm{tor} = 0.4$ is observed.

\begin{figure}[h!]
 \centering
 \includegraphics[scale=.65]{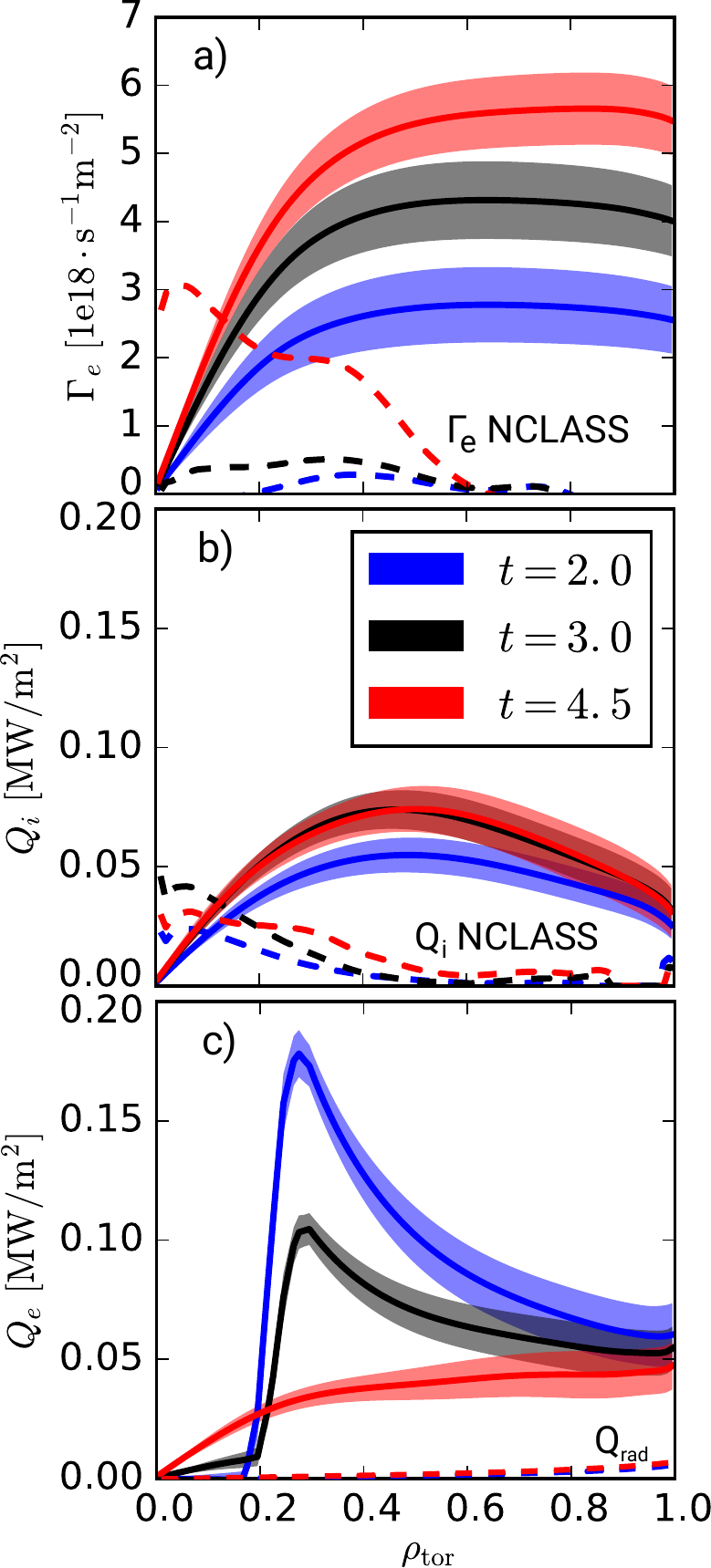}
 \caption{Particle flux from the radially integrated NBI particle source a) together with ion b) and electron c) heat flux from the TRANSP power balance. Dashed lines correspond to the neoclassical flux from NCLASS in a,b) and radiative power flux $Q_\mathrm{rad}$ in c). }
 \label{fig:175860_transp}
\end{figure}

Examination of the normalized inverse gradient scale lengths in Fig.~\ref{fig:175860_derived} defined as $R/L_X = -R_0/X\ \mathrm{d}X/\mathrm{d}r$ exposes significant changes in the transport. The region of a maximum $R/L_{n_e}$ occurs between $\rho_\mathrm{tor} = 0.4-0.7$, and it is down-shifted between the ECH and pure NBI heated cases. Given the higher NBI particle source in the latter case, this indicates a drop in an inward particle pinch or an increased diffusion. The trend is opposite inside of $\rho_\mathrm{tor} = 0.3$, suggesting a substantial change in the character of the transport.  The carbon density gradient $R/L_{n_C}$ follows nearly the same trend as $R/L_{n_e}$, profiles of $n_C$ are flatter, with twice as large a variation as $R/L_{n_e}$  close to the axis and similar difference outside. The ratio of $T_e/T_i$ drops from 1.5 to 0.65 on axis due to a reduced electron heating and the deuterium Mach number $M_D = \sqrt{m_Dv_\varphi^2/(2T_i)}$ decreases from 0.25 to 0.14 as a consequence of a lower rotation and a higher ion temperature in the NBI heated phase. 
Detailed power balance (PB) analysis and magnetic equilibrium is carried out via multiple iterations of the TRANSP suite of codes \cite{doecode12542}, and the kinetic EFIT \cite{lao1990equilibrium} constrained by \mbox{$E_r$-corrected} MSE \mbox{$q$-profiles} using the OMFIT integrated modeling framework \cite{OMFIT2015}. As is demonstrated in Fig.~\ref{fig:175860_transp}, application of 3\,MW of ECH quadruples the electron heat-flux $Q_e$ just outside of the ECH resonance, and $Q_e/Q_i$ ratio on the midradius changes from $Q_e/Q_i=$ 1.9 in the ECH heated phase to 0.5 for the pure NBI heating. Remarkably, the total power flux on a midradius is nearly constant, varying only 15\% between all phases.  The uncertainty of the fluxes is estimated from the variance within 100\,ms window. The additional ECH power increases the thermal transport coefficient for electrons $\chi_e$ by a factor of 2 and coefficient $\chi_i$ for ions by 50\% with respect to the NBI only case.  The radiated power, reconstructed from DIII-D foil bolometers\cite{leonard19952d,odstrvcil2016optimized}, is dominated by a carbon and deuterium bremsstrahlung and represents less than 5\% of the heat flux carried by electrons. The electron particle flux $\Gamma_e$ in Fig.~\ref{fig:175860_transp}a is computed as a volume integral of the particle source density deposited by NBI. Because of similar acceleration voltages and deposition profiles of all beams, $\Gamma_e$ is nearly proportional to   NBI power. The neoclassical contribution to $Q_i$ and  the Ware pinch contribution to $\Gamma_e$ is performed by the NCLASS \cite{houlberg1997bootstrap} code.  Clearly,  neoclassical ion heat flux dominates inside of $\rho_\mathrm{tor} = 0.2$,  while it is almost negligible outside of $\rho_\mathrm{tor} = 0.4$. The neoclassical pinch contributes significantly to the core particle flux only in the pure NBI-heated phase, because a lower electron temperature and thus a lower conductivity of the plasma increases toroidal electric field driving the Ware pinch\cite{wesson2011tokamaks}. The on-axis $n_e$ peaking in the last phase is likely the consequence of the Ware pinch and a higher beam fuelling. 
\vspace{-.3cm}
\subsection{Laser blow-off sytem}
\vspace{-.3cm}
Trace amounts of aluminum and tungsten are introduced in the plasma by a recently installed laser blow-off (LBO) system based on a design from Alcator C-Mod \cite{howard2011characterization}. The LBO allows multiple impurity injection,  well localized in time and with less than $\Delta t_\mathrm{source} < 1$\,ms long source function. A short source function is essential for the investigation of a fast anomalous transport because the observed impurity evolution is given by a convolution of the source function and an impurity density response on the ideal Dirac delta function source. In general, a short injection of non-recycling impurity exhibits two phases: a rapid influx often dominated by a diffusion causing a rise of plasma impurity content followed by a slow exponential decay of the impurity content and low outward flux with a nearly balanced diffusive and convective components. If the timescale of the rapid rise phase is comparable to $\Delta t_\mathrm{source}$, decoupling of diffusive and convective flux becomes difficult, and the upper boundary of the inferred diffusion would be poorly constrained. 

\subsection{Impurity diagnostics}
\vspace{-.3cm}
The lowest charge states of Al, indicating the LBO impurity source, are monitored by a fast visible camera \cite{yu2008fast} viewing the LBO port. In contrast to Al, visible radiation from W~ions is below the noise level of the camera, and no information about the edge source is thus available. 

Inside of the separatrix, the extreme ultraviolet (EUV) spectrometer SPRED \cite{fonck1982multichannel} monitors the evolution of intermediate Al charge states, namely, Al$^{10+}$ at 56.8\,nm and Al$^{8+}$ at 28.4\,nm.  The integration time of 2\,ms does not permit to resolve an abrupt rise of the impurity density after the injection, but as will be demonstrated in Sec.~\ref{sec:experimentAl}, it is a valuable constraint for the impurity recycling in the decay phase of LBO injection. Attempt to monitor tungsten trace spectroscopically, demonstrated before on DIII-D in Ref.~\onlinecite{hollmann2017core}, was unsuccessful as a result of the unavailability of short wavelength measurements and high neutron noise in the XEUS spectrometer \cite{reinke2010vacuum}. 

The SXR diagnostic\cite{hollmann2011soft} on DIII-D   serves as the primary tool constraining propagation of W~ions and is essential to monitor Al ions in the fast influx phase after injection.  The SXR system consists of two poloidally viewing cameras, placed on the outboard side symmetrically above and below midplane. Each camera observes a fan of 32 lines-of-sight (LOS), covering the whole plasma. Accurate position of LOSs and relative values of etendue are found by performing a detailed in-lab calibration. Remaining discrepancies are attributed to a variation in the effective thickness of the flat Be filters caused by a change in an incidence angle in between the central and edge lines of sight.  W injections are observed using  12.7\,$\mu$m thick Be filters to extend the radial coverage in lower temperature region at midradius. Ten-fold thicker Be filters are utilized during Al injections to reduce line radiation such that relative contributions of the fully stripped and partially stripped H-like, He-like Al ions to the total SXR radiation is nearly proportional to the density of those ions. Due to remaining systematic uncertainties, we found it impractical to infer transport coefficients directly from line-integrated brightness. Instead, a 2D tomography code \cite{odstrvcil2016optimized} is applied, providing more freedom to relax these uncertainties. Transport is thus inferred from the temporal evolution of flux surfaced averaged emissivity downsampled to 2\,kHz.  Background SXR radiation is determined from the time evolution of $n_e$, $T_e$, and $n_C$ profiles with a small offset to match average emissivity profile in 10\,ms range before injection.    

The last diagnostic constraining the evolution of fully-stripped Al ions is a CER spectroscopy. In the current experiment, 18 channels of the CER system spanning from the magnetic axis to $\rho_\mathrm{tor} = 0.85$ are dedicated to monitor $n$ = 12$\rightarrow$11 transition of an Al$^{12+}$ ion at 408.3\,nm,  produced by a charge exchange reaction between Al$^{13+}$ and the NBI neutrals. The beam is switched on continuously for 50\,ms following the LBO injections to cover the whole impurity rise and the decay phase is monitored by beam-blips. A low signal level limited the temporal resolution of CER  to  5\,ms. This resolution is sufficient to resolve a rise phase of Al$^{13+}$  density, despite a rapid Al density evolution, because of about $\sim 25$\,ms relaxation time necessary to strip He-like and H-like ions and reach the ionization equilibrium. Therefore,  CER only weakly constrains the abrupt rise of Al density dominated by He and H-like states, but it provides essential information about the radial density profile in the self-similar decay phase of the LBO injection.

Since the CER measurements are localized on the low field side (LFS) of plasma, the observed density is affected by a centrifugal asymmetry of Al ions.  To compensate for these asymmetries, we have applied a small 10-20\% correction based on the analytical approximation of asymmetry \cite{odstrcil2017physics} before comparing it to a flux surface averaged Al$^{13+}$ density evaluated by the forward model described in Sec.~\ref{sec:forward}.

The absolute value of the Al density estimated from SXR is a factor of 2--3 higher than from CER. The discrepancy is likely caused by inaccuracy in the absolute calibration or atomic data used to interpret the measurements. Because the radial flux in Eq.~\eqref{eq:ansatz} is linear in an impurity density, we can rescale the signal of both diagnostics to match density from the forward model (Sec.~\ref{sec:forward}) without effecting inferred transport coefficients.

\section{Experimental Investigation of the Impurity Transport}

\label{sec:experimentAl}

\subsection{Al injections in the ECH/NBI power scan}
The goal of this section a thorough investigation of the laser blow-off (LBO) injection and deriving the profiles of the transport coefficient, which will be later compared with gyrokinetic modeling.
The aluminum transport is examined in the discharge \#175861, identical to \#175860 discussed in the Sec.~\ref{sec:setup} except that a majority of CER spectrometers are tuned to the Al line.
The Al particles are injected by LBO into each heating phase at 2.0\,s, 3.0\,s, and 4.5\,s. The delay between the LBO is arranged to be about eight-fold the impurity confinement times $\tau_\mathrm{imp}$, to minimize overlap between decays phases of the injections.  Non-perturbative trace behavior of the impurity is verified by an electron temperature measurement from ECE diagnostic and a fast density reflectometer, where both indicate less than 5\% perturbation inside of the pedestal. Further, $Z_\mathrm{eff}$ inferred from visible bremsstrahlung increase by a mere 5\%, and the impurity concentration did not exceed 100\,ppm. None of these LBOs triggered ELM, and the first ELM did not occur earlier than 25\,ms after injection, providing enough time for Al ions to penetrate over the low diffusion region in  ETB into the plasma core. The impurity confinement time $\tau_{imp}$, determined from the decay rate of CER signal on-axis, gradually increases in between heating phases from 120\,ms in the high ECH case to 160\,ms in the low ECH case and up to 300\,ms in the NBI heated phase, while the energy confinement time $\tau_E$ improves from 115\,ms to 130\,ms, then up to 150\,ms respectively. The ratio $\tau_{imp}/\tau_E$ thus doubles from 1.0~to~2.0. 
\begin{figure}[h]
 \includegraphics[scale=.7]{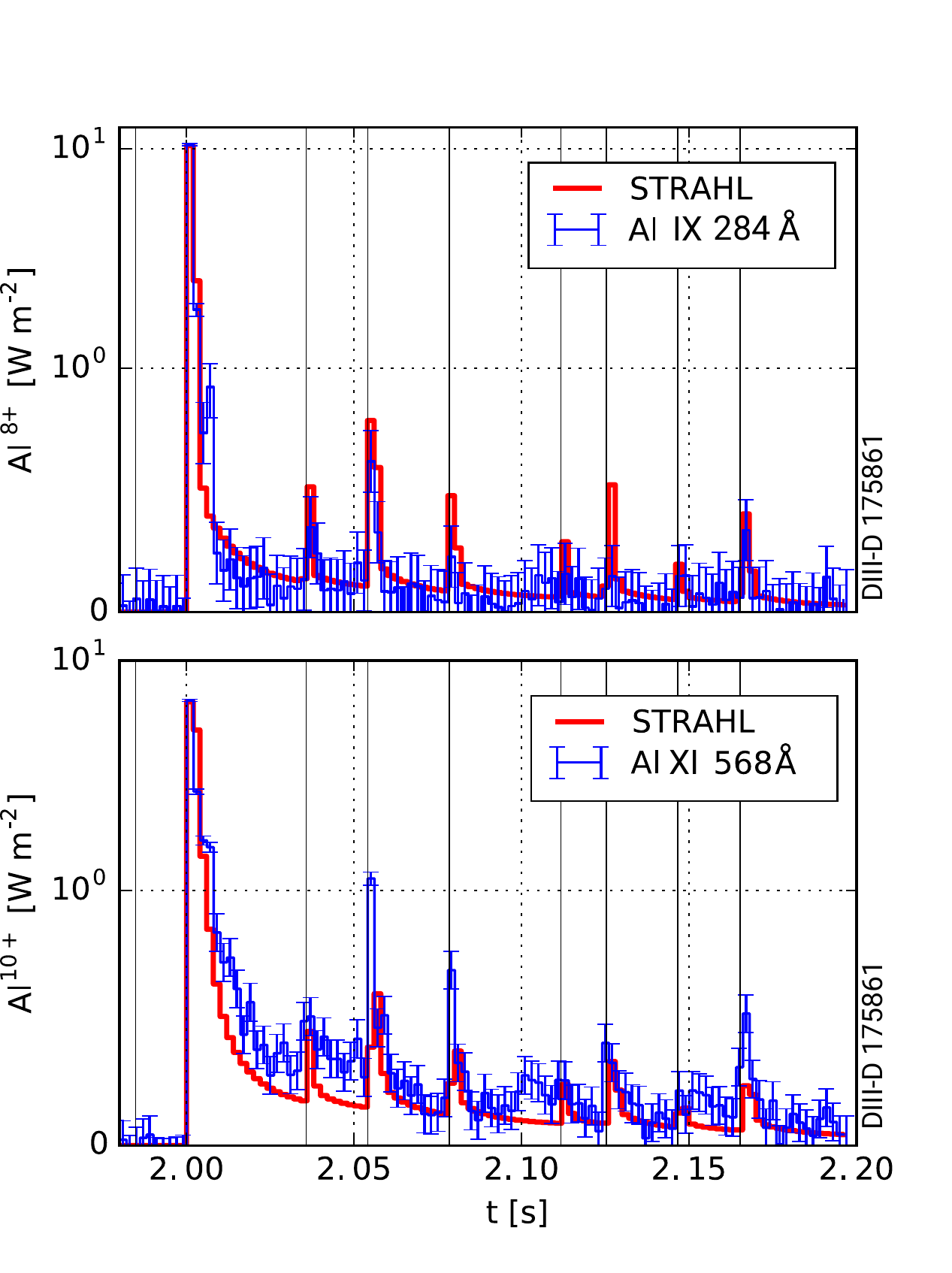}
 \caption{Brightness of two VUV Al lines (Al$^{8+}$ 284\AA{}, Al$^{10+}$ 568\AA{}) observed by SPRED spectrometer after the LBO injection are shown by a blue line. The solid red line corresponds to a signal calculated maximum from a posteriori estimate of the transport coefficient via STRAHL, and thin vertical lines indicate the timing of the ELMs. }
 \label{fig:retro_spred}
\end{figure}

Evolution of  Li-like and B-like aluminum observed by SPRED in Fig.~\ref{fig:retro_spred} provides information about particle source, because  the recombination rate of these ions is so low that without a source of a neutral impurity particles these ions are not present in the plasma. 
In less than 2\,ms after ablation at 2.0\,s, the observed radiation reaches its maximum, and then the brightness rapidly declines as the impurity propagates towards the core and particles are ionized to higher charge states. The nonzero level of Al$^{10+}$ signal after the fast decay phase, as well as the spikes in Al$^{10+}$ and Al$^{8+}$  following the ELMs can be explained only if  Al ions are partially recycling from walls\cite{victor2019pedestal}.

\begin{figure}
 \centering
 \includegraphics[scale=.7]{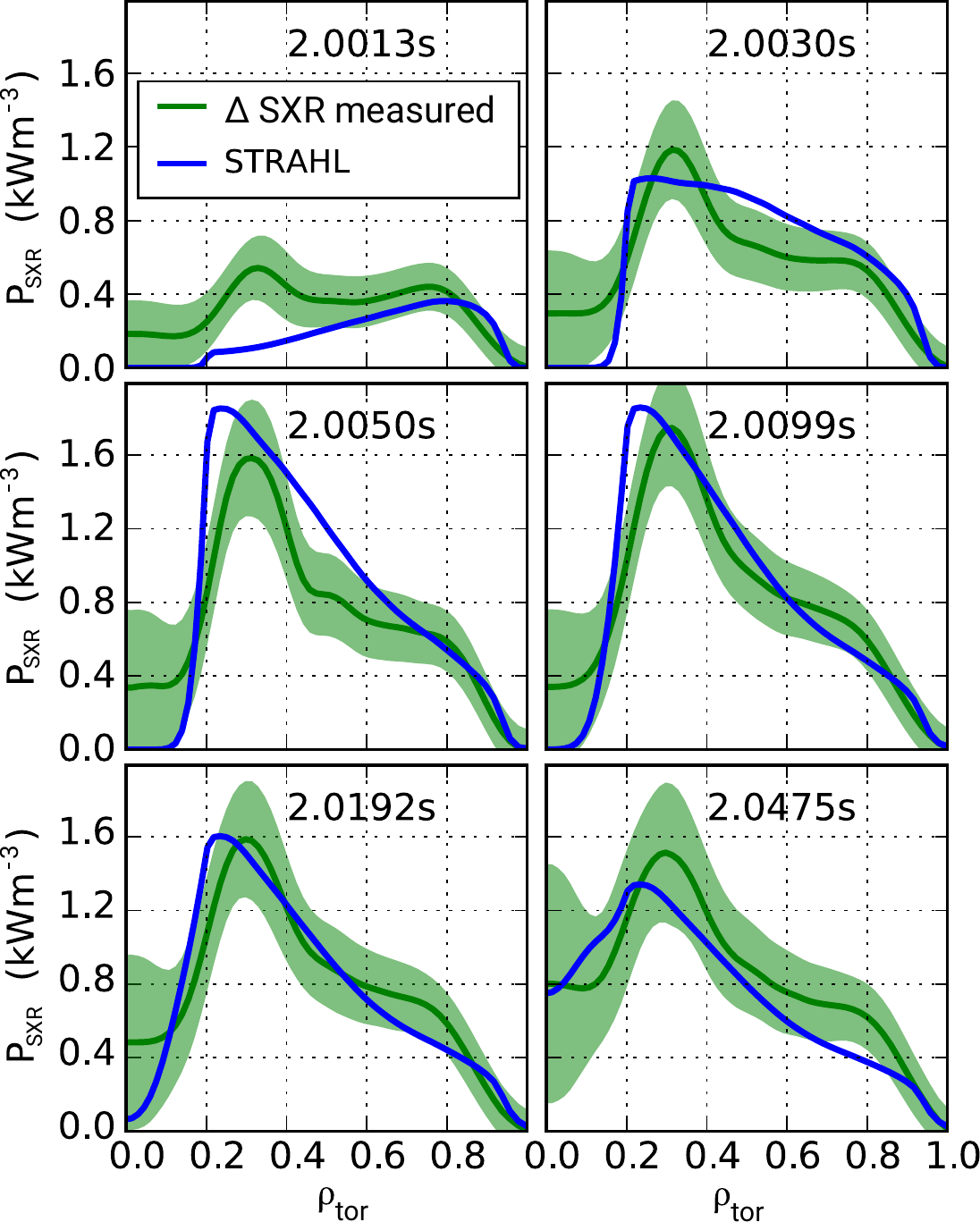}
 \caption{Evolution of the background subtracted local soft X-ray emissivity from the tomographic inversion following the first LBO injection of Al at 2.0\,s.   }
 \label{fig:retro_SXR}
\end{figure}

SXR emissivity of Al constrains a rapid density influx phase and provides essential information for decoupling of transport coefficients. This illustrates the SXR emissivity following the first ablation in 2.0\,s at Fig.~\ref{fig:retro_SXR}. In the region outside of $\rho_\mathrm{tor} = 0.25$, the   SXR emissivity equilibrates in less than 10\,ms, and later it decays self-similarly,  indicating an unusually fast transport. In contrast, the evolution is particularly slow inside of $\rho_\mathrm{tor} = 0.25$  and  the emissivities do not equilibrate until 50\,ms after the injection.  
 The  SXR signal is dominated by He-like, H-like, and fully-stripped Al ions,  and the emissivity closely follows the evolution of actual impurity density. SXR emissivity thus provides the essential information constraining the Al evolution during the fast influx phase, which CER observing only Al$^{13+}$ cannot deliver.  However, the data are utilized only up to the first occurrence of ELM because of difficulties with proper background subtraction. 

\begin{figure}
 \centering
 \includegraphics[scale=.7]{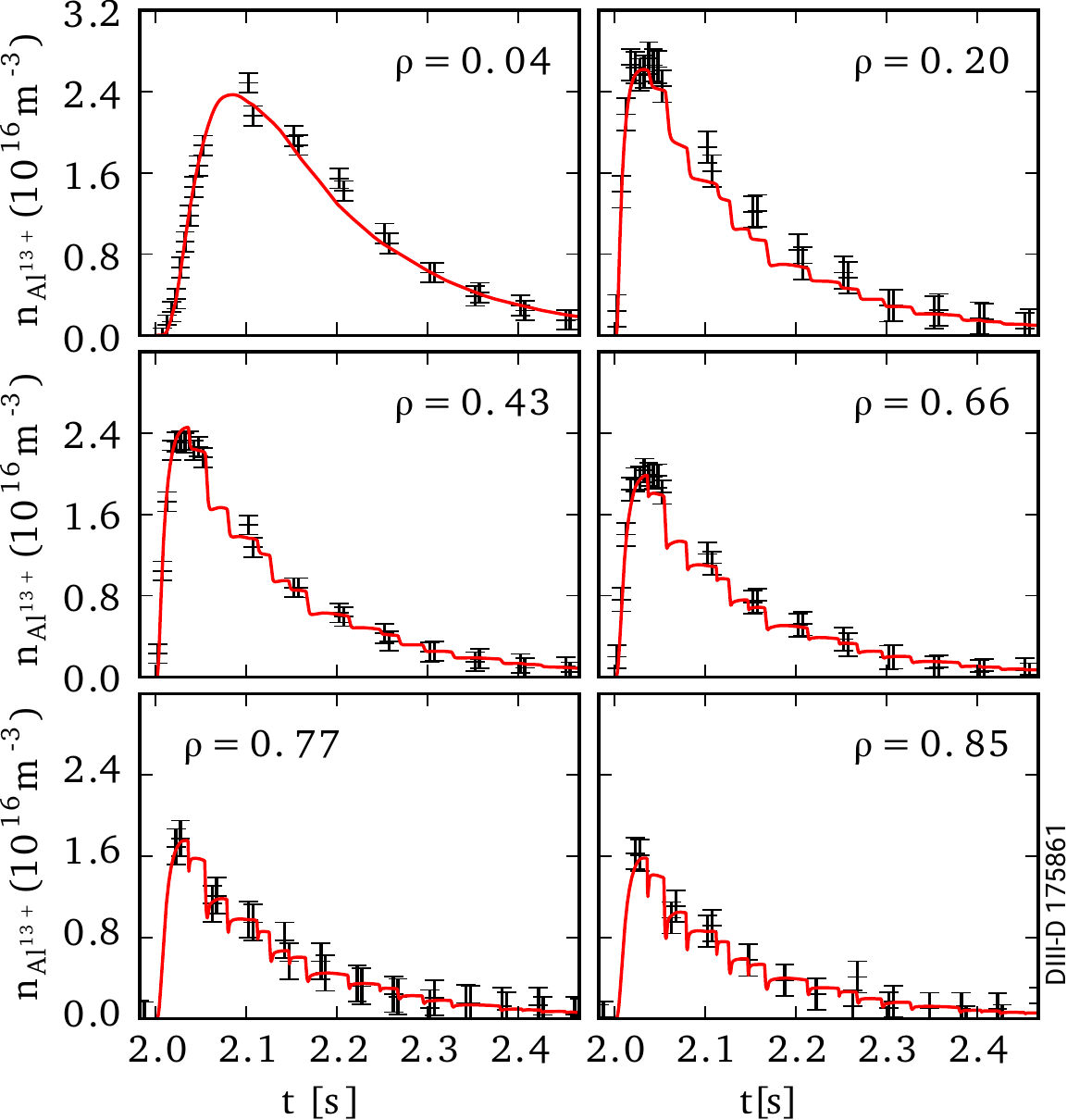}
 \caption{ Temporal evolution of Al$^{13+}$ density after LBO in 2.0\,s measured by CER (black errorbars) and matched forward model (red line). Stair-like decay of signal results from a drops of impurity pedestal during each ELM and consecutive inward propagation.}
 \label{fig:retrofit_cer_time}
\end{figure}

\begin{figure}
 \centering
 \includegraphics[scale=.7]{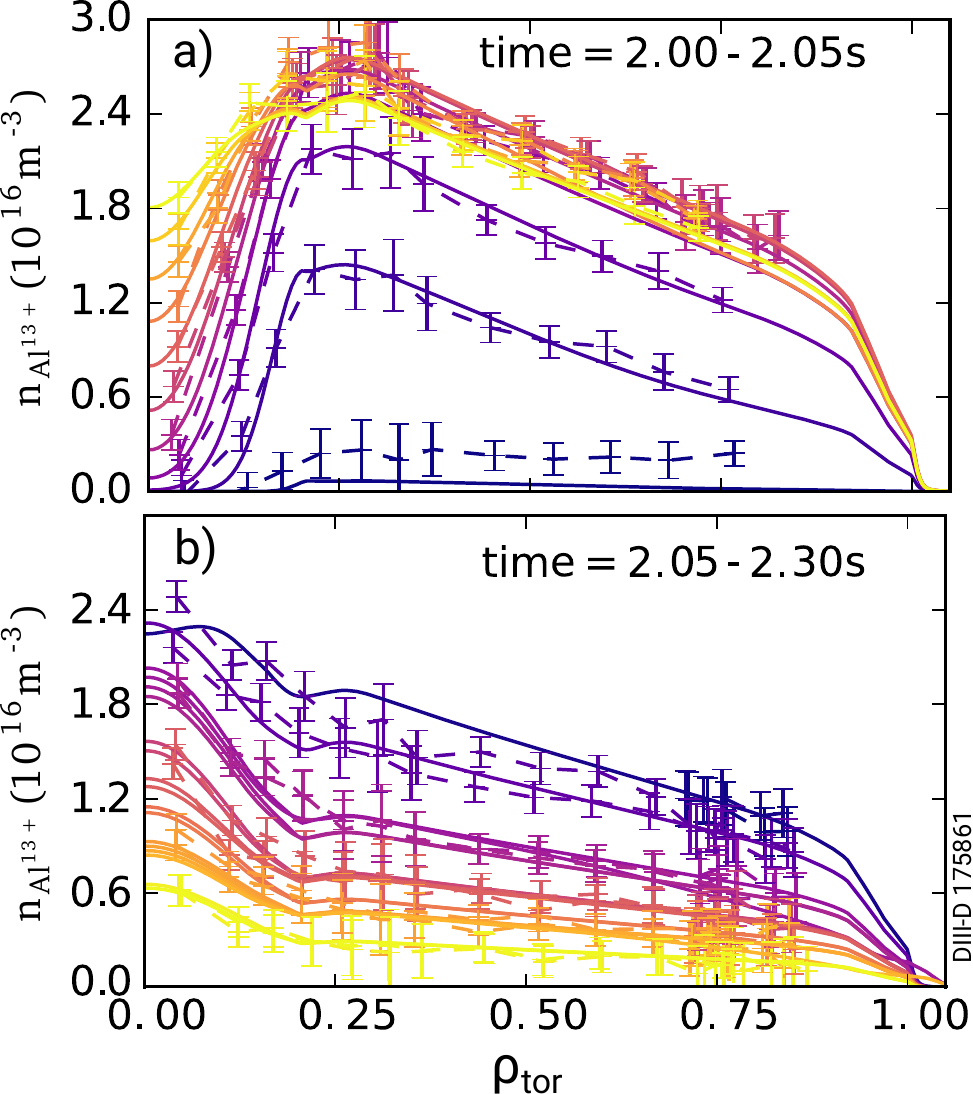}
 \caption{ Radial profiles of Al$^{13+}$ density measured by CER after LBO at 2.0\,s, a) in a rise phase, b) in a decay phase. The solid lines indicates a density calculated by the forward model. }
 \label{fig:retrofit_cer}
\end{figure}

About 10-20\,ms after the injection, the fully-stripped ions start to dominate the fractional abundance of Al, and the CER signal begins to rise swiftly, as shown in Fig.~\ref{fig:retrofit_cer_time}. 
The observed evolution of Al$^{13+}$ is tightly followed by a forward model, matching both the rise and the decay phase of LBO injection.   Radial profiles  from Fig.~\ref{fig:retrofit_cer} confirms a presence of slow transport region inside of $\rho_\mathrm{tor} = 0.25$. As a consequence of the low diffusion, the density profile of Al$^{13+}$ is initially hollow on-axis,   while later, it peaks in the decay phase. Peaking is also a  result of a higher fractional abundance of fully-stripped Al ions increasing from 85\% at $\rho_\mathrm{tor} = 0.3$ up to 98\% on-axis. ELMs produce rapid drops in pedestal impurity density (see Fig.~\ref{fig:retrofit_cer_time}), propagating through a fast transport domain on midradius up to the region of slow diffusion in the vicinity of a magnetic axis. ELMs appear to be responsible for a major fraction of impurity flow over ETB,  switching them off in the forward model triples the impurity confinement time $\tau_{imp}$ to 360\,ms in the ECH  case, while $\tau_{imp}$  remains constant in the NBI case.

\begin{figure*}
 \centering
 \includegraphics[scale=.8]{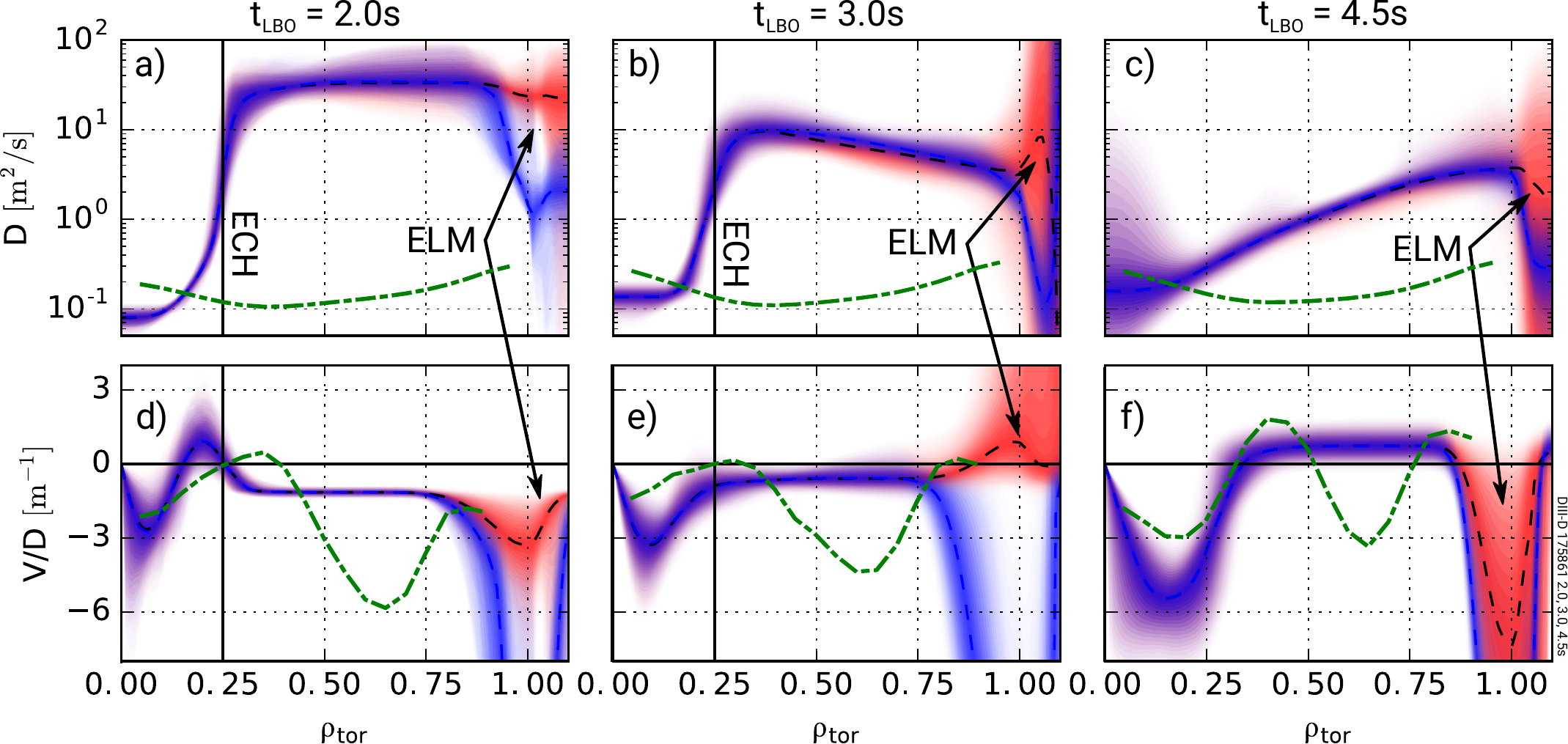}
 \caption{Inferred probability distribution of Al diffusion $D$ a-c) and normalized convection $V\!/D$ d-f) for high ECH  a,d), low ECH  b,e) and pure NBI case c,e). Blue contours indicates inter-ELM profiles, while red contours corresponds to a profile observed during ELMs. Green dashed-dotted line is a neoclassical value from NEO code. }
 \label{fig:175861_transport}
\end{figure*}

\begin{figure}
 \centering
 \includegraphics[scale=0.7]{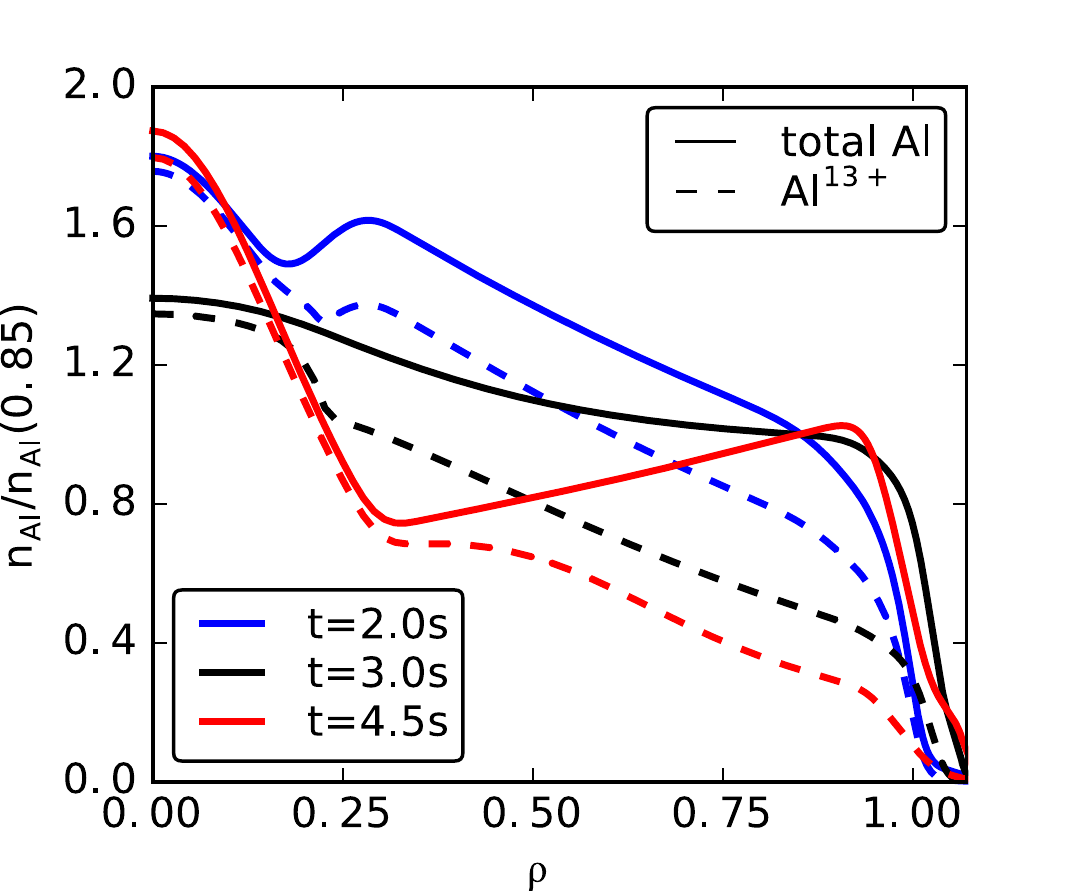}
 \caption{Predicted stationary density profiles of aluminum normalized to the value at outermost CER measurement at $\rho_\mathrm{tor} = 0.85$ (full line) and density of fully-stripped Al shown by the dashed line.  }
 \label{fig:ion_abundance}
\end{figure}

The massive increase in the infered transport coefficients in Fig.~\ref{fig:175861_transport} reflects 
the large change of the observed impurity propagation.   Inside of ECH resonance $\rho_\mathrm{tor} = 0.25$ the diffusion $D$ is particularly low, about $D = 0.1\mathrm{\,m^2/s}$ and close to the neoclassical value from the drift-kinetic code NEO\cite{belli2008kinetic}. In the same region, $V\!/D$ becomes more negative as the ECH power diminishes. Since $-V\!/D$ is equal to the normalized gradient of stationary impurity density $\nabla n_z / n_z$, a more negative value of $V\!/D$ indicates a significant increase of the on-axis peaking. $V\!/D$ is about zero in the 3\,MW ECH heated case, while in the NBI only condition, $V\!/D$ drops to $-5$\,m$^{-1}$, implying on-axis accumulation of Al ions. The neoclassical $V\!/D$ inside of $\rho = 0.25$ roughly reproduces this observation. The core accumulation is likely a consequence of a higher particle source from NBI and the Ware pinch leading to on-axis peaking of the bulk ions that increases $\nabla n/n$ driven neoclassical inward pinch.
Just outside of the ECH resonance occurs a marked increase in the impurity diffusion, with the inferred diffusion increasing by more than two orders of magnitude compared to $D_{neo}$. Diffusion reaches a value of up to about $D \sim 30\mathrm{\,m^2/s}$ in the high ECH case, and to $ 10\mathrm{\,m^2/s}$ in the low ECH case. 
In contrast, the diffusion in pure NBI case grows steadily only up to $D \sim 1\,\mathrm{\,m^2/s}$ at midradius. The inferred stationary impurity profile outside of ECH resonance (Fig.~\ref{fig:ion_abundance}) are slightly peaked with $V\!/D \sim-1$\,m$^{-1}$ in high ECH case,  nearly flat with $V\!/D \sim-0.5\,$m$^{-1}$ and slightly hollow $V\!/D \sim 0.5$\,m$^{-1}$ in the pure NBI case. Note that a similar trend is present also in stationary carbon profiles (see Fig.~\ref{fig:175860_derived}).  
The radial profiles are constrained mostly by CER measurements of fully-stripped Al ions, which is the lowest in the NBI case (see Fig.~\ref{fig:ion_abundance}) because of low $T_e$ and slowest impurity transport. Therefore, we must point out that uncertainty in recombination and ionization rates unaccounted in the uncertainty can affect $V\!/D$ values inferred in the NBI case. 

Since the outermost available CER channel measuring Al is located inside of the pedestal at $\rho_\mathrm{tor} = 0.85$, the information about pedestal transport is inferred only indirectly from the SPRED VUV lines, and particle conservation constrained by the observed evolution impurity density inside and measured source at SOL.  Inferred pedestal transport also depends on our parameterization of $D$, $V\!/D$ profiles and ELMs in the forward model. Despite these difficulties, some profile features are very robust. The impurity diffusion in pedestal decreases by order of magnitude to $D = 0.1-1\mathrm{\,m^2/s}$ and $V\!/D$ becomes strongly negative with a minimum of order of $-100$ and a considerable uncertainty (not shown), in agreement with previous studies\cite{puiatti2003simulation, putterich2011elm, dux2003chapter,valisa2011metal}.  ELMs are reproduced by an increase of ETB $D$ to values comparable with core $D$, and a drop in ETB $V\!/D$  to nearly zero.

\subsection{W~injections in the ECH/NBI power scan}
\label{sec:w_transport}
The trace amount of tungsten is introduced to plasma and contrasted with Al injections to investigate a charge and mass dependence of the impurity transport. Each W LBO ablated about $\sim 3\cdot 10^{17}$ particles into the discharge \#175886 with the identical heating scan as in \#175860 discussed in Sec.~\ref{sec:setup}. The impurity propagation was monitored by the fast SXR diagnostic utilizing a thin 12.7\,$\mu$m thick Be filter. Despite the reduced thickness of this filter, the measured W signal decreased below the noise level outside of the midradius.  The absence of any local information about the edge W~density prevented meaningful inference of the pedestal transport profiles. Therefore, we have adopted the approach from Ref.~\onlinecite{dux1999z} where the transport equation for total density was integrated inward from the outermost reliable measurement, and  STRAHL is applied only for an iterative adjustment of the SXR cooling factor $L_W^\mathrm{SXR} \equiv \varepsilon_{SXR}/(n_Wn_e)$.

\begin{figure}
 \centering
 \includegraphics[scale=0.5]{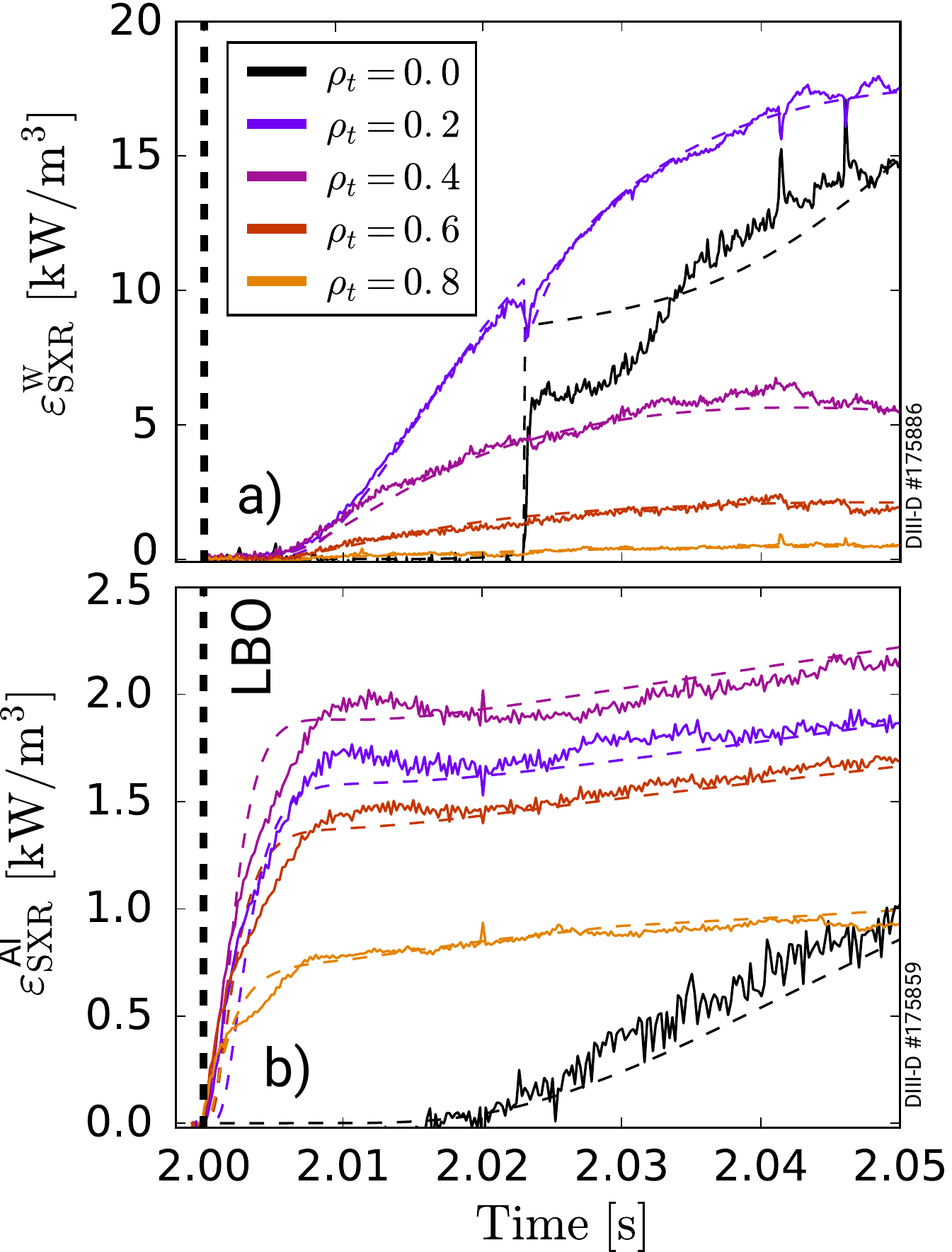}
 \caption{Time evolution of background subtracted SXR signals for a) W~ and b) Al ablation at 2.0\,s. The dashed line is a fit from the forward model described in Sec.~\ref{sec:w_transport} and on-axis discontinuity in W~case at 2.023\,s is caused by a sawtooth crash.  }
 \label{fig:sxr_evolve}
\end{figure}
\begin{figure*}
 \centering
 \includegraphics[scale=0.45]{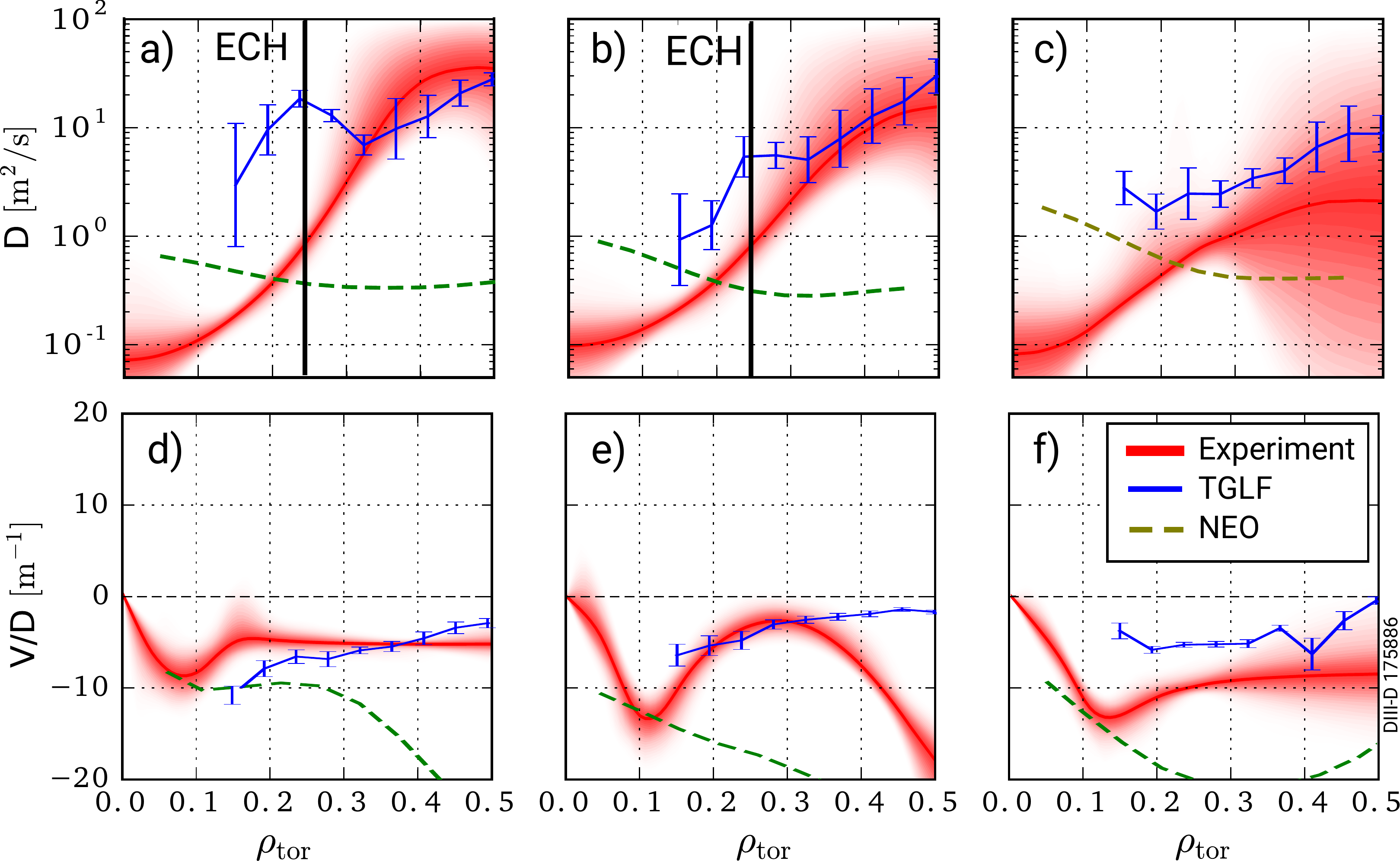}
 \caption{ The red contours corresponds to inferred profiles of W~diffusion a-c) and normalized pinch d-f) for high ECH a,d), low ECH b,e) and only NBI case c,f).  The blue line with errorbars comes from $Q_i$ and $Q_e$ heat flux matched TGLF model\cite{staebler2007theory} and green dashed line is a value from NEO code. The potato orbit radius\cite{dux2003influence} of W~ions is $\rho_\mathrm{tor} = 0.04$.}
 \label{fig:175886_DV_Ws3}
\end{figure*}

The propagation of Al and W during a raise phase after LBO in Fig.~\ref{fig:sxr_evolve} illustrates a remarkable difference in transport between these impurities.  While Al emissivity equilibrates in 10\,ms, W~needs at least 50\,ms, indicating a significantly lower transport either at the edge or in the plasma core. The impurity confinement time $\tau_{imp}$ of tungsten increased from 350\,ms in the first ECH heated phase up to 1200\,ms in the pure NBI heated phase, which is 3-4$\times$ longer than $\tau_{imp}$ for aluminum. The actual transport coefficients inferred inside of $\rho_\mathrm{tor} = 0.5$ are plotted in Fig.~\ref{fig:175886_DV_Ws3}. The W~diffusion on-axis is about $D=0.1$\,m$^2$/s independently of ECH heating, in agreement with the observation for aluminum. Outside of ECH resonance, the diffusion increases above 10\,m$^2$/s in the full ECH case, and about 1\,m$^2$/s in pure NBI case. Given the significant uncertainty in the inferences, these profiles are in quantitative agreement with Al in Fig.~\ref{fig:175861_transport}, only the increase at the ECH location is less sharp,  likely due to a lower spatial resolution of the SXR tomography compared to the local CER measurements. This result is notable as it implies that there is not a strong Z dependence of the impurity transport in regions of the plasma dominated by turbulence (outside of the ECH deposition).  The diffusion coefficients from the TGLF model\cite{staebler2007theory} roughly reproduces the magnitude, but not the observed trend in the scan. The difference in $D$ and $V$ between Al and W calculated by TGLF is less than 30\%. 
$V\!/D$ inside of $\rho_{\mathrm{tor}} = 0.3$ becomes more negative with declining ECH power, resulting in a minor on-axis accumulation in the NBI only case. 
Outside of this region, the $T_e$ dependence of the tungsten cooling factor plays a dominant role in determining the inferred transport.  Therefore, despite smaller statistical uncertainties, the inferred $V\!/D$ is likely to be dominated by inaccuracies associated with the complex atomic physics of radiation of W ions.

Neoclassical diffusion $D_\mathrm{neo}$ calculated by the NEO code and shown in Fig.~\ref{fig:175886_DV_Ws3} by a green dashed line, exceeds experimental values by about an order of magnitude near the axis. We note that the agreement of NEO in this region with the inferred transport of Al was quite good (see Fig.~\ref{fig:175861_transport}).  However, Al is considerably less massive (A$_{Al}$ = 27.0\,AMU) and therefore is not as susceptible to centrifugal asymmetries.   The large mass of W (A$_W$ = 183.8\,AMU) leads to a significant LFS impurity accumulation, which substantially increases the magnitude of neoclassical transport\cite{angioni2014neoclassical, belli2014pfirsch}. Without including the centrifugal asymmetry effects in NEO,  on-axis $D_\mathrm{neo}$ is about 0.05\,m$^2$/s, below the experimental value.  The current setup of SXR cameras does not allow us to measure the poloidal asymmetry of W~with a necessary precision to verify the centrifugal force model, but recent experiments on AUG\cite{odstrcil2017physics} revealed a significant impact of fast ions in low-density discharges. Nonetheless, including the fast ions in the poloidal force balance reduced $D_\mathrm{neo}$ by only 20\%, insufficiently to explain the observed difference. Investigation of this discrepancy will be a subject of the future study on DIII-D.

\section{Interpretative gyrokinetics simulations of Al transport} 
\label{sec:gyro}
Experimental transport coefficients derived in the previous sections are now contrasted with state of the art gyrokinetic (GK)  simulations performed by  CGYRO\cite{candy2016high, belli2017implications} code. Our main objective are the experimental validation of the gyrokinetics predictions in dominant ion and electron heated plasmas. CGYRO is an Eulerian GK solver designed for collisional, electromagnetic, and multi-species simulations with sonic rotation capability\cite{belli2018impact}. The computation cost of the simulations is reduced by performing ion scale GK runs, including only electrostatic fluctuations. This should be sufficient to capture the ion temperature gradient (ITG) and trapped electron modes (TEM), the main contributors of particle and heat transport in the core of typical DIII-D H-modes. The simulations include the effects of rotation ($E\!\times\! B$ shear, etc.) and collisions are accounted by Lorentz collisional operator\cite{belli2017implications}, suitable for a low collisionality ($ \bar{\nu}_e < 0.04$) observed in our experiment.  The flux surface geometry is described by a Miller equilibrium\cite{miller1998noncircular}, providing a sufficiently accurate parameterization in the core of moderately shaped plasmas. 

The GK simulations include four ion species - bulk deuterium ions, carbon, and two trace aluminum ions with a different density gradient and $n_{Al}/n_e = 10^{-5}$. Because the impurity in a trace limit does not affect quasi-neutrality,  the gradient flux relation \eqref{eq:ansatz} remains linear\cite{angioni2009gyrokinetic, skyman2012impurity} and the impurity flux is unambiguously described by $D$ and $V$ coefficients~from~Eq.~\eqref{eq:DVconverion}.

\subsection{Nonlinear CGYRO modeling}
\label{sec:nonlin}

\begin{table}
\centering
\caption{Parameters of the nonlinear CGYRO simulations: number of radial $n_r$ and toroidal $n_{tor}$ modes, poloidal grid points $n_\theta$,  box size $n_\mathrm{box}$, and size of simulation domain $L_x/\rho_s\times L_y/\rho_s$. Maximum wavenumber $k_x$ was set close to $10\rho_s$ and $k_y$ to about 1$\rho_s$. }
\smallskip 
\label{tab:cgyro}
{\renewcommand{\arraystretch}{1.2}
\begin{ruledtabular}
\begin{tabular}{@{}lccccccc@{}}
Case             & $n_\mathrm{r}$&  $n_{tor}$     & $n_\theta$    &n$_\mathrm{box}$     & $L_x/\rho_s$     & $L_y/\rho_s$\\\hline
2.0\,s, $\rho = 0.43$    & 288        &      18    &16        & 6    &108    &98\\
3.0\,s, $\rho = 0.43$    & 300        &      16    &16        & 6     &71    &74\\ 
4.5\,s, $\rho = 0.43$    & 400        &      16    &24        & 8     &118     &98\\ 
2.0\,s, $\rho = 0.62$    & 300        &      18    &24        & 10     &108     &104\\ 
4.5\,s, $\rho = 0.62$    & 300        &      18    &16        & 10     &92     &104\\ 
\end{tabular}
\end{ruledtabular}
}
\end{table} 

We have performed a set of five nonlinear  simulations  that investigate the turbulence in each of the heating mixes (ECH/NBI) at two locations $r/a = 0.5$ ($\rho_\mathrm{tor}  = 0.43$) and $r/a = 0.7$ ($\rho_\mathrm{tor} = 0.62$). For the low ECH case, analysis at $r/a = 0.7$ was not included because of issues with CGYRO convergence. The resolution parameters for each case are summarized in Tab.~\ref{tab:cgyro}. A  convergence test performed  for the first case included 50\% increase of all resolution parameters one by one, without any significant effect on the computed fluxes.
\begin{figure}
 \centering
 \includegraphics[scale=.7]{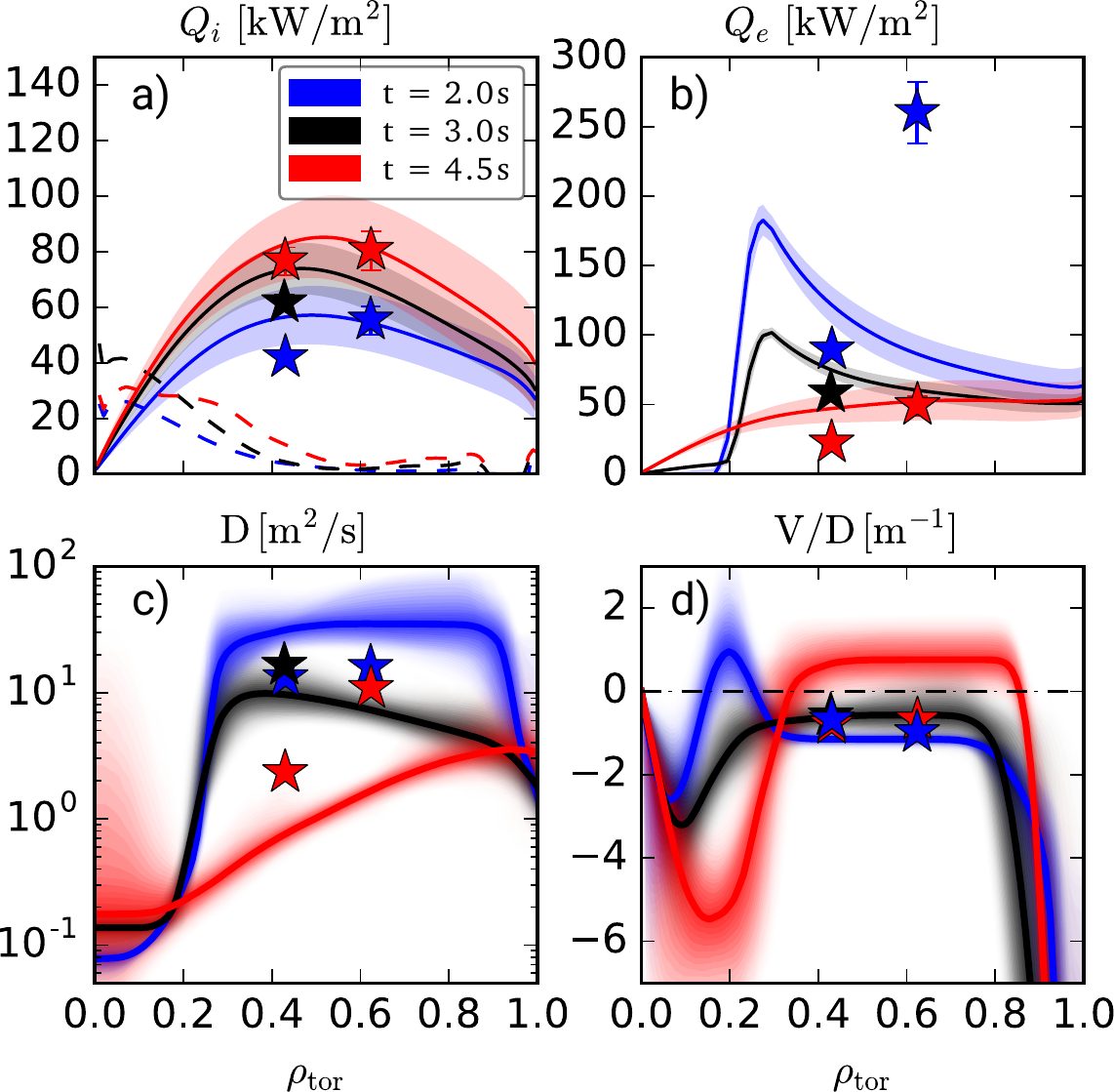}
 \caption{Comparison of experimental radial profiles (full lines with confidence intervals) with nonlinear and neoclassical results summed together (stars) for a) ion heat flux, b) electron heat flux, c) Al diffusion, d) Al normalized convection $V\!/D$.  }
 \label{fig:exp_prof_nonlincgyroA}
\end{figure}

Uncertainties in experimental gradients and the strong dependence of the GK model on these gradients results in large uncertainty in the calculated GK fluxes. Since it is assumed that the ion heat flux arises from the ion-scale turbulence captured in the simulations (while electron heat flux can arise from a variety of scales), the  $Q_i^{GK}$ of each case was matched with anomalous ion heat flux,  defined as $Q_\mathrm{anom} = Q_i^{PB}-Q_\mathrm{neo}$. This is done by varying $R/L_{T_i}$ within $2\sigma$ uncertainty, while gradients of $T_e$ and $n_e$ were fixed on experimental values. Due to stiff transport, the calculated particle and heat flux varied nearly by order of magnitude.  Figure~\ref{fig:exp_prof_nonlincgyroA} shows the experimental profiles compared to the calculated heat fluxes and transport coefficients that resulted from heat flux matching these conditions. As a result of the ion heat flux matching, $Q_i^{GK}$ from CGYRO summed with $Q_i^{NEO}$ overlaps with $Q_i^{PB}$. In the case of electron heat flux $Q_e^{GK}$, experimental flux is above gyrokinetic in two ECH heated cases at  $\rho_\mathrm{tor} = 0.43$ which could be a result of missing high-k TEM/ETG turbulence contribution or uncertainties in gradients not varied during this scan. The high ECH case at $\rho_\mathrm{tor} = 0.62$ is deep in the TEM regime, and it was not possible to reduced $Q_e/Q_i$ ratio without also varying electron temperature and density gradients.  Calculated Al diffusion in Fig.~\ref{fig:exp_prof_nonlincgyroA}c shows a substantially lower variation than observed experimentally. Diffusion is factor of 2-3 below experimental value in high ECH case, close in low ECH case and factor 4 and 10 above the diffusion in NBI heated case at $\rho_\mathrm{tor} = 0.43$ and $\rho_\mathrm{tor} = 0.62$. 

The ratio between the convective and diffusive fluxes  $V\!/D$ is only weakly dependent on the saturated fluctuation level and thus less sensitive to the experimental uncertainties than $D$ proportional to the flux magnitude. The direction of the convection predicted by gyrokinetic simulation is inward, about $V\!/D \sim -1$ in all five cases, close to a value inferred in the ECH-heated cases. However, in the pure NBI heated case, the outward convection was observed, and despite the uncertainties discussed in Sec.~\ref{sec:experimentAl}, it is implausible that the actual $V\!/D$ is equal to ECH heated cases.  The gradient of a directly measured Al$^{13+}$ density is nearly the same in all cases (see Fig.~\ref{fig:ion_abundance}) and the drop in a fractional abundance of Al$^{13+}$ ion due to a reduced $T_e$ would inevitably lead to a flatter total Al density in the NBI phase and since  $1/L_{n_{Al}} = -V\!/D$ also to more positive $V\!/D$. A similar range of values is found for a gradient of carbon density (Fig.~\ref{fig:175860_derived}) with a negative  $1/L_{n_C}$ (positive $V\!/D$) in the NBI case outside the midradius. 

\begin{figure}[h]
 \centering
 \includegraphics[scale=.63]{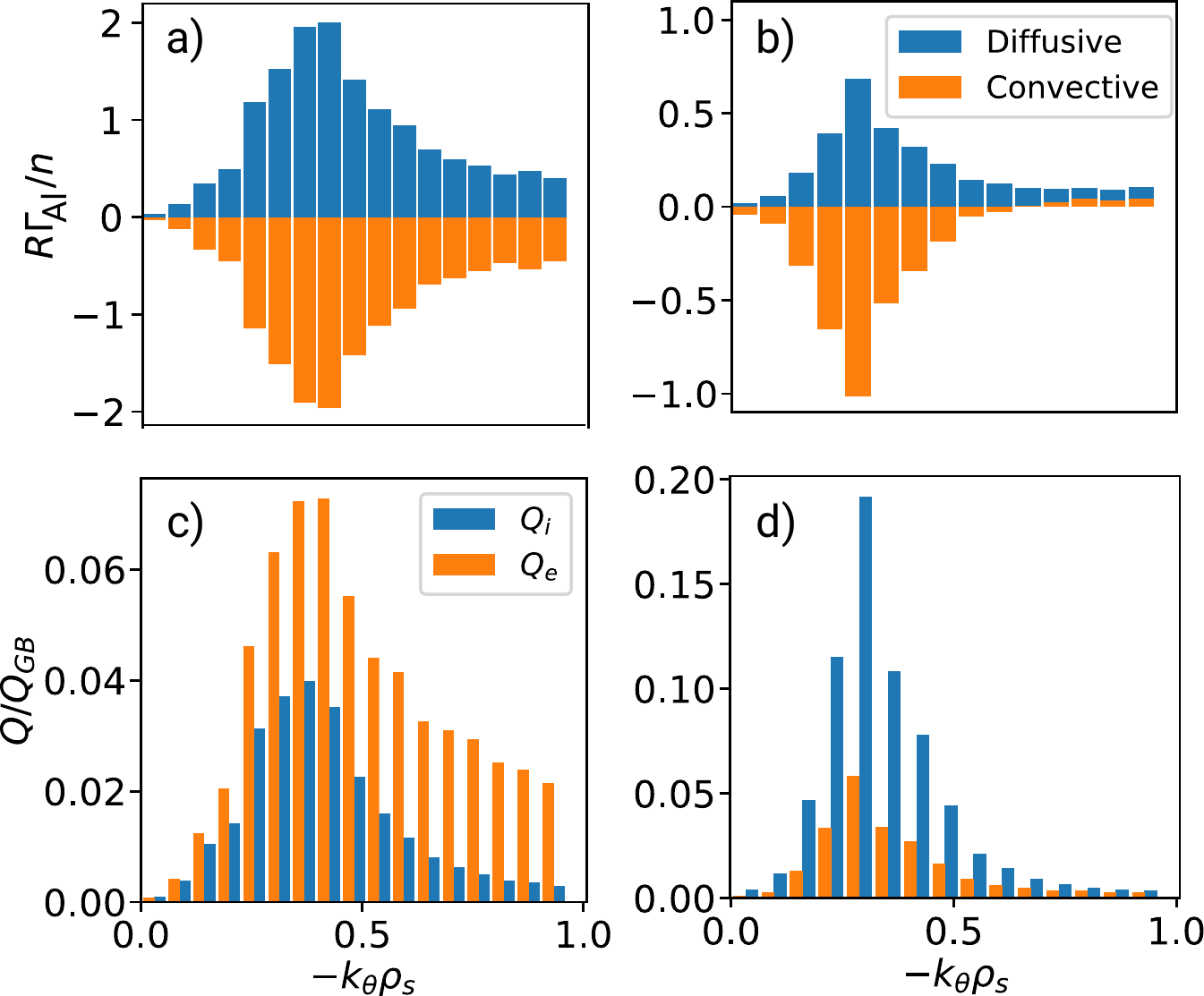}
 \caption{Spectra of normalized convective and diffusive Al flux a,b) and heat flux c,d) for the ECH  case a,c) and the NBI heated case b,d) at $\rho = 0.43$. Trace impurity gradient used to calculate the diffusive flux is defined to zero the total impurity flux.  }
 \label{fig:5_DV_nonlinear}
\end{figure}

Spectra of heat and particle flux in Fig.~\ref{fig:5_DV_nonlinear} helps to identify optimal binormal wave number $k_y\rho_s$ for linear simulations shown later as well is to investigate truncating of the flux at maximum $k_y/rho_s$ value. 
 Both particle and heat flux spectra for the ECH case in Fig.~\ref{fig:5_DV_nonlinear}a,c) are broader and shifted towards higher wavenumbers than in the pure NBI case at Fig.~\ref{fig:5_DV_nonlinear}b,d).
The electron heat flux in the ECH case is truncated at the highest wavenumber, and this may indicate an important role of intermediate or high-$k$ modes in setting the electron heat flux $Q_e^{GK}$.  The particle flux is truncated as well, but it unlikely explains a factor of 3 underestimated experimental diffusion. On the contrary, the NBI case is well resolved in the ion scale $k_y$ range.  Convective and diffusive flux in  Fig.~\ref{fig:5_DV_nonlinear}a,b) have both a similar $k_y$ dependence, with a maximum for $k_y\rho_s = 0.45$ and  $k_y\rho_s = 0.35$ in ECH and NBI case, respectively.

\begin{figure}[h]
 \centering
 \includegraphics[scale=0.35]{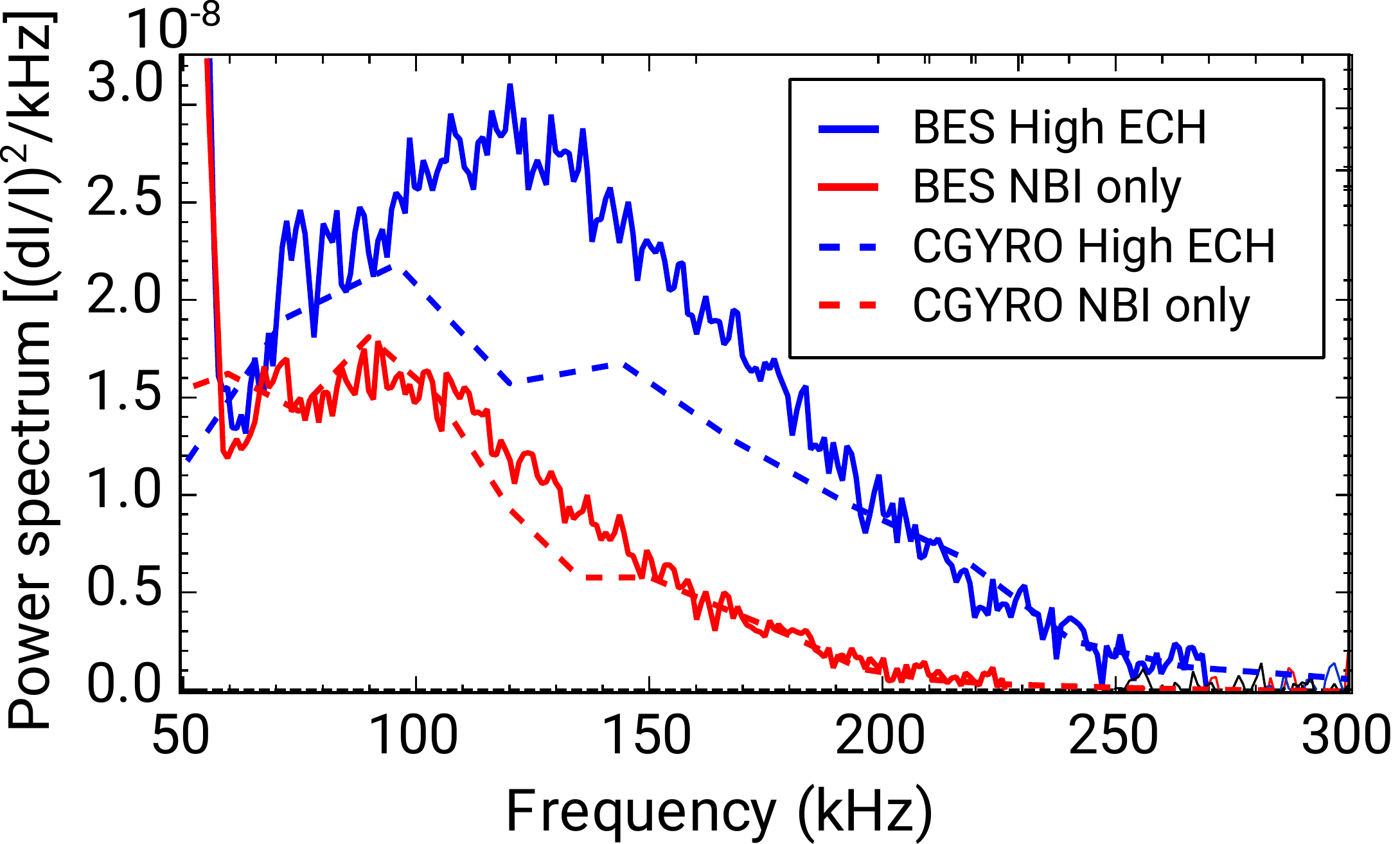}
 \caption{Power spectrum of intensity fluctuations observed by BES diagnostic at $\rho_\mathrm{tor} = 0.63$. Extend of the BES point-spread function allows to observe only long-wavelength turbulence with $k_y\rho_i < 1$. }
 \label{fig:BES_spectrum}
\end{figure}

The validation of CGYRO simulations is finished by comparing computed density fluctuations with a beam emission spectroscopy (BES)\cite{mckee1999beam} observing D$_\alpha$ emission of the beam neutrals colliding with background plasma. The fluctuation intensity of this radiation is proportional to the variation in a local electron density $\delta n_e$. The synthetic diagnostic used for this analysis is adapted from previous GYRO simulation work \cite{holland2009implementation}, where the time evolution of the density fluctuations $\delta n_e/n_e$ are obtained from a non-linear CGYRO run, and Fourier transformed into the real space. 
The synthetic signals are computed by integrating over a BES point spread function and dividing by a factor of 2 to account for atomic physic processes related to the beam emission\cite{holland2009implementation}.
The synthetic power spectrum in Fig.~\ref{fig:BES_spectrum} is found to be in a satisfactory agreement with the measured BES spectra. This is true for both the shape of the spectrum (dominated by the Doppler shift) and the fluctuation magnitude. These measurements thus independently validate fluctuation level determined from matching of an experimental ion heat flux. It is important to note that the agreement with BES is not always consistent with good agreement in other fluxes (as shown in Fig.~\ref{fig:exp_prof_nonlincgyroA}). This emphasizes the need for multi-channel validation and comparisons at all levels of the primacy hierarchy \cite{greenwald2010verification}.

\subsection{Quasi-linear CGYRO modeling}
\label{sec:quasilin}
Possibilities for the interpretative modeling of the experiment by nonlinear simulations are limited by their huge computational cost. Therefore, to investigate the role of dominant turbulent modes, we have adopted the approach developed in Ref.~\onlinecite{angioni2015gyrokinetic} and apply a quasi-linear method based on CGYRO simulations at a single representative wave number  $k_y\rho_s = 0.4$. The choice of this wavenumber is motivated by a position of maxima of heat and particle fluxes in our nonlinear simulations.  Since the absolute magnitude of fluxes depends on an unknown nonlinearly saturated fluctuation amplitude, particle and heat fluxes must be quantified by a ratios, for example $Q_e/Q_i$ and $D/\chi_\mathrm{eff}$  where effective heat conductivity $\chi_\mathrm{eff}$ is defined as $\chi_\mathrm{eff} = (Q_i+Q_e)/(n_i\mathrm{d}T_i/\mathrm{d}r+n_e\mathrm{d}T_i/\mathrm{d}r)$. Maximizing the ratio $D/\chi_\mathrm{eff}$ is equivalent to identifying conditions that maximize turbulent impurity diffusion while simultaneously minimize degradation of energy confinement.

\begin{figure}[h]
 \centering
 \includegraphics[scale=.7]{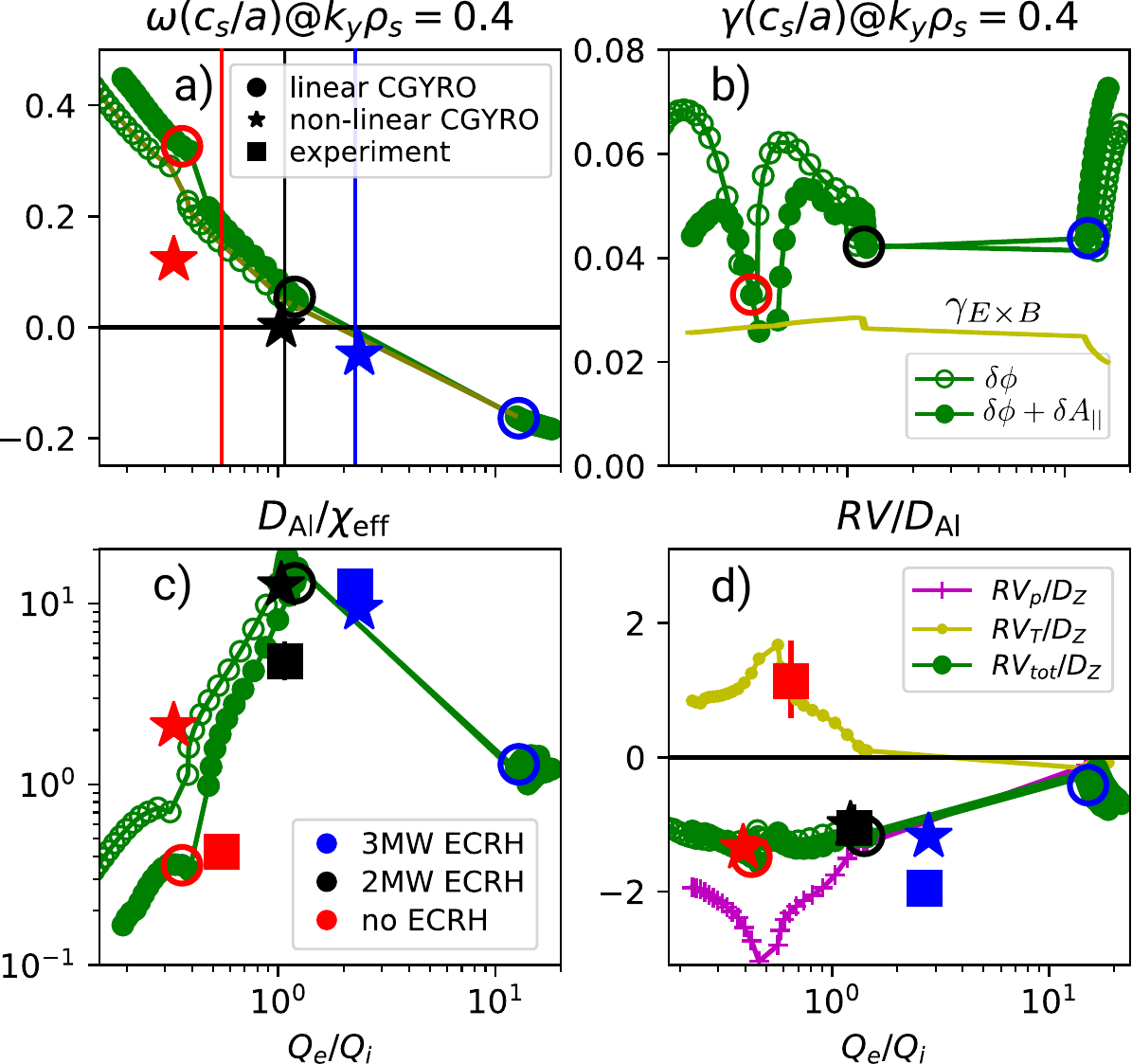}
 \caption{Real eigenfrequency a), linear growth rate with $E\times B$ shearing rate $\gamma_{E\times B} = -(r/q)\partial \omega_0/\partial r$  for reference b), $D_\mathrm{Al}/\chi_\mathrm{eff}$ ratio c) and $R(V\!/D)_\mathrm{Al}$ ratio d) for linear results plotted as function of electron to ion heat flux ratio $Q_e/Q_i$ at $\rho_\mathrm{tor} = 0.43$.  Full circles corresponds to   linear simulations including $\delta \phi$ and $\delta A_{||}$ fluctuations, open includes only $\delta \phi$, nonlinear results are shown by stars and experimental values by squares. Vertical lines in a) corresponds to experimental values of anomalous $Q_e/Q_i$. Convection in d) is decomposed in a thermodiffusion component $RV_T\!/D_\mathrm{Al}$ (small circles) and pure convection $RV_p\!/D_\mathrm{Al}$ (crosses).}
 \label{fig:cgyro_linear}
\end{figure}

The quasilinear modeling in Fig.~\ref{fig:cgyro_linear} was performed not only for the three heating steps at $\rho_\mathrm{tor} = 0.43$ but also for values interpolated along a linear trajectory in the parameter space and extrapolated outside the measured range. These simulations are compared with the results of nonlinear GK runs and experimental measurements. In Fig.~\ref{fig:cgyro_linear}a, the real eigenfrequency $\omega$ of the mode at $k_y\rho_s = 0.4$ decreases monotonically with $Q_e/Q_i$ ratio from a positive value of the real frequency (in ion drift direction) for the ITG dominated NBI case, to negative (in electron drift direction)  in  TEM dominated high ECH case. The discontinuity between ECH  cases is a consequence of the ITG/TEM transition. In fact,  the turbulent state close to the transition is composed of a mixture of both modes. And as the additional ECH increases $\nabla T_e$, more TEM is destabilized, exhausting the additional  $Q_e$.  Since the TEM leads to particularly stiff transport in the electron channel, the 50\% increase in ECH input power leads to a mere 10\% increase in the measured value of $R/L_{T_e}$ on midradius (see Fig.~\ref{fig:175860_derived}a). 

Time averaged mode frequency in the nonlinear simulation is down-shifted towards zero with respect to the linear results by the contribution of the linearly subdominant modes. Also, $Q_e/Q_i$ ratio in high ECH case is reduced by additional ITG driven $Q_i$ flux.
Still, the relation between the frequency $\omega$ and $\log(Q_i/Q_e)$ remains nearly linear and $Q_i/Q_e$ thus serves as a macroscopic proxy for $\omega$.  The linear grow rate $\gamma$  in  Fig.~\ref{fig:cgyro_linear}b varies only moderately between three examined cases and $\gamma$ in all cases is above $E\times B$ shearing rate $\gamma_{E\times B}$. Including electromagnetic fluctuations ITG reduced grow rate  and value of $D/\chi$ due to a finite $\beta_e$ effects\cite{hein2010electromagnetic}. Since the electromagnetic fluctuations are not included in the nonlinear CGYRO runs, the missing $\beta_e$ stabilization may be responsible for higher values of $D/\chi$ and $Q_i/Q_e$ observed in the modeled NBI heated case.  
\begin{figure}
 \centering
 \includegraphics[scale=0.6]{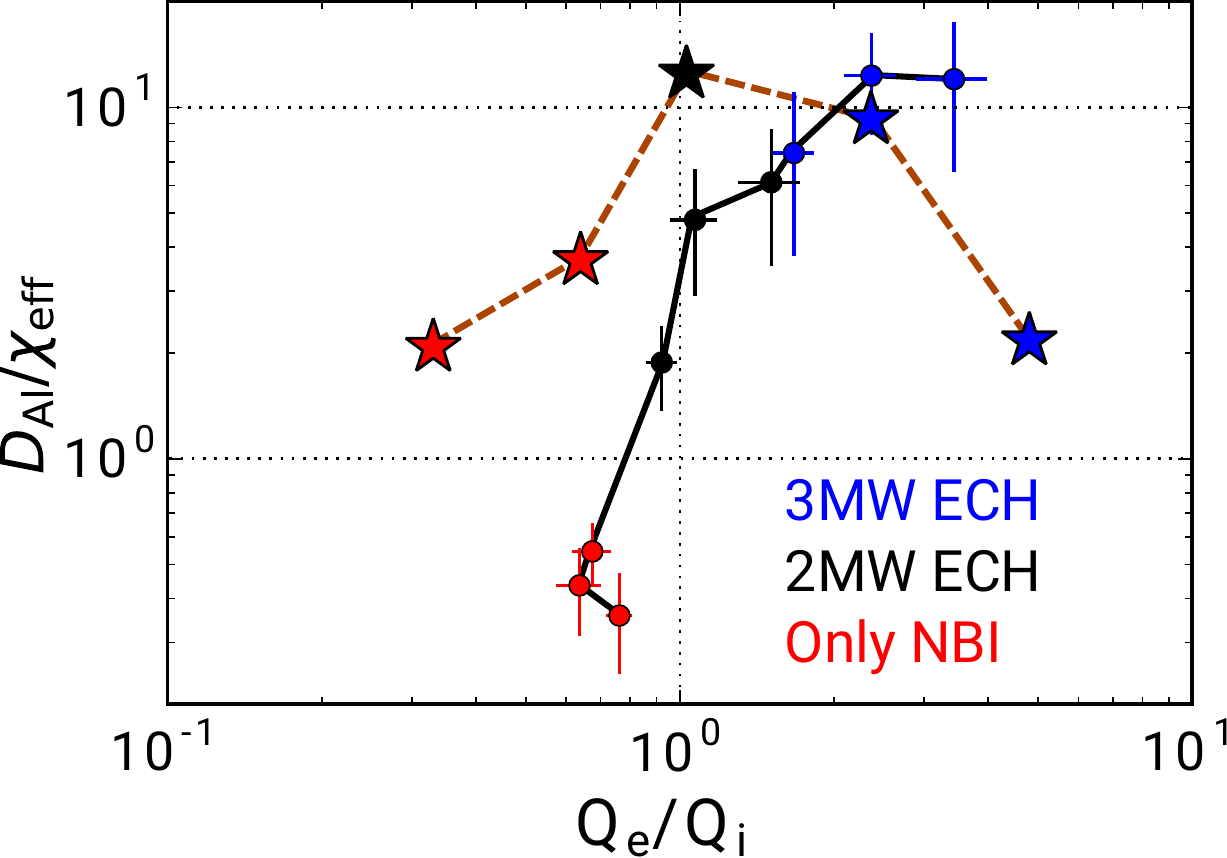}
 \caption{Nonlinear results from CGYRO code plotted by stars are contrasted with experimental }
 \label{fig:DchiQeQi_scan}
\end{figure}

The next step is to assess the role of dominant and subdominant modes in impurity diffusion.  In Fig.~\ref{fig:cgyro_linear}c both, linear and nonlinear simulations of $D/\chi_\mathrm{eff}$,  peak for $Q_e/Q_i\sim 1$ as the real frequency of the mode approaches zero. This result is in agreement with recent GKW and analytic model results described in Ref.~\onlinecite{angioni2015gyrokinetic}.  The linear results, including $\delta A_{||}$ fluctuations, reproduce the experimental $D/\chi_\mathrm{eff}$ ratio rather well, except for high ECH case in TEM regime. 
Since the nonlinear simulation for the high ECH case is in agreement with the experiment, it is speculated that this is a result of an unaccounted contribution from subdominant modes. 
On the contrary, the nonlinear simulation performs worse in ITG dominated NBI case, yielding a factor of 5 higher $D/\chi_\mathrm{eff}$ than measured. The reason could be attributed to the overestimated contribution of linearly subdominant modes propagating in electron drift direction together with unaccounted $\delta A_{||}$ fluctuations. Assuming that  $\omega$ and it's proxy $Q_e/Q_i$ are the main parameter determining the $D/\chi_\mathrm{eff}$ ratio enables us to plot in Fig.~\ref{fig:DchiQeQi_scan} the nonlinear results obtained from all radii as a function of a common variable. While $D/\chi_\mathrm{eff}$ ratio from  nonlinear CGYRO  peaks for $Q_e/Q_i\sim 1$, experimental ratio only saturates for $Q_e/Q_i \gg 1$ without any clear decline. Moreover, the experimental measurements indicate a much steeper rise when $Q_e$ approaches $Q_i$, and we can conclude that less additional electron heating is required to enhance impurity diffusion than is predicted by nonlinear simulations. 

A proper understanding the impurity peaking is as important as to the diffusion, and therefore the measured Al peaking factor $RV\!/D$ is contrasted with GK modeling in Fig.~\ref{fig:cgyro_linear}d.  Linear results for the convection are decomposed in the thermodiffusion $V_T$ - proportional to temperature gradient of the impurity and pure convection $V_p$, which consists of parallel compression term, curvature pinch, and rotodiffusion\cite{angioni2012analytic,dubuit2007fluid, camenen2009impact}. Thermodiffusion leads to outward convection of impurities for modes in ion drift direction and small inward convection for modes in electron direction. The pure pinch is dominated by the inward parallel compression proportional to $q$. A variation in $V_T$ and $V_p$ nearly cancel each other in the ITG dominated conditions while the two contributions add in conditions with dominant TEM. The total convection is thus inward and about $RV\!/D\sim -1$ for ITG  and 0.5 for the TEM-dominated case.   Moreover, in both ITG-dominated cases, quasilinear estimate matches exactly the nonlinear results, while the TEM-dominated high ECH case is mismatched due to a missing contribution from the subdominant ITG modes. In contrast to diffusion, including of $\delta A_{||}$  fluctuations does not affect $RV\!/D$. Experimental value increases from  $RV\!/D = -1.9\pm 0.2$ in the high ECH case, over  $RV\!/D = -1.0\pm 0.3$ in the low ECH case up to $RV\!/D = 1.1\pm 0.5$ in the NBI heated case. Variation in $R/L_{T_i},R/L_{T_e},R/L_{n_e},T_e/T_i,q, \hat{s}$ and $\nu^*$  well outside of an expected uncertainty did not revealed any parameter which could significantly increase outward convection or reduce $D/\chi_\mathrm{eff}$ in quasilinear simulation to reverse a sign of $RV\!/D$ or match experimental $D/\chi_\mathrm{eff}$  ratio.

\section{Conclusions}
\label{sec:end}

In this paper, we have presented a dedicated impurity transport study on the DIII-D tokamak designed to perform a multi-channel validation of the gyrokinetic simulation. For this purpose,  a standard ELMy H-mode scenario was optimized using predict-first simulations of the core plasma to maximize variation in impurity flux, while maintaining optimal conditions for DIII-D‘s impurity diagnostic access. 
The impurity transport was probed by trace LBO injections of aluminum in all of the conditions studied. To our knowledge,  this is the first attempt to combine LBO, a nearly ideal impurity source,  with high-resolution and local  CER measurements, SXR emissivity, and edge VUV lines to constrain impurity transport inferences. The outstanding quality of impurity measurements, in combination with a detailed forward model based on STRAHL, were essential for constraining the impurity transport. The distribution function for transport coefficients was estimated via Bayesian inference, which is believed to deliver more statistically rigorous fits and uncertainty estimates than a standard $\chi^2$-minimization method.

The experiments represented a scan of the heating mix (NBI/ECH) performed with an approximately fixed total heat flux at midradius and constant torque.
A factor of 4 variation in electron to ion heat flux ration $Q_e/Q_i$ and factor of 2 variation in $T_e/T_i$ on midradius were achieved by a partial decoupling of electron and ions in low collisionality plasmas. Other main parameters like $n_e$, $\omega_\varphi$, $Z_\mathrm{eff}$ and magnetics equilibrium remained approximately constant. 
In the NBI heated phase, low values of $T_e/T_i $ and steep $T_i$ gradients  destabilized ITG modes.  This is in contrast to the ECH dominated phases of the discharge where the low collisionality, high $T_e/T_i$ ratio,  $R/L_{n_e}$ and  $R/L_{T_e}$ created favorable conditions for the occurrence of TEM.  
The application of electron heating caused a remarkable increase of impurity diffusion just outside of ECH resonance. Diffusion at midradius increased by a factor of 30, and the impurity confinement time $\tau_{imp}$ decreased 2.5$\times$ with respect to the pure NBI case. The lower  $\tau_{imp}$ is achieved despite a factor of 2 reduction in the ELM frequency.  This occurs as a consequence of the larger ELMs and substantially faster core diffusion, both leading to enhanced particle flush during the ELM crash. Since $Z_\mathrm{eff}$ remains constant, ELMs must also increase the influx of intrinsic impurities to balance the losses.  An on-axis peaking of impurity density is observed in the NBI case as a consequence of a higher particle source, Ware pinch, and low particle diffusion on-axis.  In these conditions, the increased ion density gradient drives an inward neoclassical pinch of impurities. This is in contrast to the reduced Ware pinch and particle source in the ECH heated case that prevents impurity accumulation even inside of the heating radius. 

Aluminum transport is compared with tungsten transport by W LBO injection in the identical discharge conditions, revealing significant differences in the propagation of these impurities. The rise phase following the injection as well as the impurity confinement time are several times longer for W than Al.  Since the inferred core diffusions are similar, it is likely a consequence of a slower pedestal and edge tungsten transport. In contrast to aluminum, W ions exhibit on-axis peaking even in the presence of ECH, and the peaking of both impurities substantially increased when the electron heating is replaced by NBI. The experimental $V\!/D$ in the inner core is qualitatively explained by neoclassical modeling.   However, a significant mismatch of the experimental and neoclassical W diffusion is not well explained by the poloidal asymmetry model implemented in the NEO.

The second part of the paper presented interpretative gyrokinetic modeling of aluminum transport. Nonlinear, ion-scale gyrokinetic simulations, constrained by experimental heat flux, have shown an order of magnitude increase in impurity diffusion, in quantitative agreement with the experiment. The most significant discrepancy was found for the NBI heated phase at $\rho_\mathrm{tor} = 0.63$, where CGYRO predicted a diffusion that exceeded experimentally inferred values $10$-times. This discrepancy is remarkable since the electron and ion heat flux, as well as the BES fluctuations at this radius, are well-matched.  In the ECH-heated case at $\rho_\mathrm{tor} = 0.63$, the electron heat channel was overestimated by a factor of 3, despite matching ion heat flux channel and BES fluctuations.

Additional investigation was performed using linear gyrokinetics simulation of a single representative wave number. We have confirmed the results of the previous theoretical study\cite{angioni2015gyrokinetic}, claiming that the ratio of impurity diffusion to thermal diffusivity is maximized for modes with eigenfrequency shifted apart from the values which maximize ion and electron heat fluxes respectively. $D/\chi_\mathrm{eff}$ thus peaks around $\omega \sim 0$, while  $Q_e/Q_i$ decreases monotonously with eigenfrequency $\omega$.  Due to a strong dependence of $D/\chi_\mathrm{eff}$ on mode frequency $\omega$ and large difference between TEM and ITG, the impurity transport is particularly sensitive to a nonlinearly saturated fraction of these modes. Variation of the other CGYRO input parameters, even when changed well outside of the expected uncertainty range, does not affect the properties of the dominant mode enough to explain observed discrepancies.  
Overall, the dependence of $D/\chi_\mathrm{eff}$ on  $Q_e/Q_i$  is significantly steeper in the experiment, leading to a conclusion that a small increment of $Q_e$ in  $Q_e<Q_i$ regime enhances impurity diffusion substantially more than shows ion scale electrostatic gyrokinetics simulation.     

Aluminum peaking gradually increased from $RV\!/D \sim -2$ in the ECH phase to $\sim 1$ in the NBI phase, and follows a trend observed in electron and carbon density.  This trend disagrees with gyrokinetics simulation that predicts a value $RV\!/D \sim -1.5$  for all three heating phases. A similar experimental trend was observed in a study of intrinsic low-Z impurity density profiles\cite{angioni2011gyrokinetic,casson2013validation}, revealing hollow boron profiles in NBI heated discharges, while when  ECH has been introduced the profiles has peaked. Whereas the latter was well reproduced by a gyrokinetics code, hollow profiles were not.  The discrepancy was correlated with a gradient of the plasma rotation, however a limited variation on $\omega$ and its gradient  do support  the same conclusion in our experiment. The aluminum $V/D$ trend is opposite to a prediction published in a recent nonlinear gyrokinetics study\cite{angioni2016gyrokinetic} for high-Z impurities showing a significant increase of outward convection for $Q_e > Q_i$ dominated by pure convection $V_p$, partially canceled by an inward thermodiffusion~$V_T$.

Results in this paper confirm the beneficial effect of an additional electron heating in reducing the impurity confinement and avoidance of high-Z impurity accumulation.  An increase in electron heat flux in ITG dominated regime excite TEM modes and significantly increase impurity diffusion outside of the heating radius in quantitative agreement with gyrokinetic codes.  Faster core diffusion then results in higher inter-ELM impurity loss and significantly lower impurity confinement time, despite a reduced ELM frequency and inward midradius convection. 
Electron heating, provided either by alpha particles or externally, together with an efficient transport mechanism over ETB, are thus essential for maintaining a low impurity content in future fusion devices. 

In this context, further work is required to investigate impurity transport in L-mode, where it can be achieved a lower collisionality and more TEM dominated plasmas. Moreover, the role of pedestal transport needs to be examined more.  Despite an order of magnitude increase in core impurity diffusion, only a factor of 2.5 drop in $\tau_\mathrm{imp}$ was observed, attributed to the rise of the inter-ELM transport. Pedestal transport in between the ELMs seems to be weakly affected by a core electron heating.   


\begin{acknowledgments}
The authors would like to thank T. P{\"u}tterich for providing atomic data necessary for the interpretation of soft-X ray signals. This material is based upon work supported by the U.S. Department of Energy, Office of Science, Office of Fusion Energy Sciences, using the DIII-D National Fusion Facility, a DOE Office of Science user facility, under Award(s) DESC0014264 and DE-FC02-04ER54698, and uses resources of the National Energy Research Scientific Computing Center (NERSC), a U.S. Department of Energy Office of Science User Facility operated under Contract No. DE-AC02-05CH11231 for nonlinear gyrokinetics simulations. Linear gyrokinetics modeling and Bayesian inference presented in this paper were performed on the MIT-PSFC partition of the Engaging cluster at the MGHPCC facility (www.mghpcc.org) which was funded by DoE grant number DE-FG02-91-ER54109. Part of the data analysis was performed using the OMFIT integrated modeling framework \cite{OMFIT2015}.

 \end{acknowledgments}

\section*{Data availability}
The data that support the findings of this study are available from the corresponding author
upon reasonable request.

 \section*{Disclaimer}
This report was prepared as an account of work sponsored by an agency of the United States Government. Neither the United States Government nor any agency thereof, nor any of their employees, makes any warranty, express or implied, or assumes any legal liability or responsibility for the accuracy, completeness, or usefulness of any information, apparatus, product, or process disclosed, or represents that its use would not infringe privately owned rights. Reference herein to any specific commercial product, process, or service by trade name, trademark, manufacturer, or otherwise does not necessarily constitute or imply its endorsement, recommendation, or favoring by the United States Government or any agency thereof. The views and opinions of authors expressed herein do not necessarily state or reflect those of the United States Government or any agency thereof.
\appendix

\section{ Forward model for the impurity density evolution}
\label{sec:forward}
 
Transport analysis was performed via the impurity transport code STRAHL \cite{dux2014strahl}, which solves a system of coupled continuity equations for each ionization stage. By choosing a radial coordinate $r = \sqrt{V_r/(2\pi^2R_0)}$, where $V_r$ is volume enclosed by a flux surface, and after flux surface averaging of all quantities, the continuity equations can be written in cylindrical geometry even for non-circular plasma cross section: 
\begin{equation}
 \frac{\partial n_Z}{\partial t} -\frac{1}{r}\frac{\partial}{\partial}r\Gamma_z = \left[ -\left(S_{Z}+\alpha_Z\right)n_{Z} +S_{Z}n_{Z-1}+\alpha_{Z}n_{Z+1} \right]n_e
 \label{eq:continuity}
\end{equation} 
where $n_Z$ is a flux surfaced averaged density of ionization stage $Z$, $\Gamma_Z$ is a radial flux of this stage and  $S_{Z}$ and $\alpha_{Z}$ are ionization and recombination rate coefficients.  
For the impurity density in a trace limit, the particle flux is assumed to be a sum of the diffusive and convective term:
\begin{equation}
 \Gamma_z = -D \frac{\partial n_Z}{\partial r}+Vn_Z,
 \label{eq:ansatz}
\end{equation} 
where $D(r)$ and $V(r)$ are diffusion coefficient and drift velocity, respectively. We assume that $D$ and $V$ are the same for all ions at a given radius, which is justifiable since only a narrow range of ionization stages occupy each radial location and the weak charge dependence of $D$ and $V$ found in previous studies \cite{guirlet2009anomalous,giroud2007method}. 

An alternative definition of the transport coefficient commonly adopted in transport codes is to derive $D_0$ and $V_0$ with respect LFS density $n_0$ and minor radius coordinate $r_m$. In order to compare with the inferred experimental coefficients, the following conversion is performed: 
\begin{align}
  D &=  D_0 \tilde{n}\left(\frac{\partial r}{\partial r_m} \right)^{\!\!\!2} \nonumber  \\
V &=  V_0\tilde{n}\frac{\partial r}{\partial r_m}    +D \frac{1}{\tilde{n}}\frac{\partial\tilde{n}}{\partial r_m} 
\label{eq:DVconverion}
\end{align}
 where $\tilde{n}$ stands for poloidal asymmetry factor $\tilde{n}  \equiv n_0/\langle n \rangle$. Detailed formulas, derived as a function of local quantities and their gradients, can be found in the appendix of Ref.~\onlinecite{angioni2014tungsten}. A difference in between these definitions becomes apparent for asymmetric poloidal profiles of heavy impurities like tungsten, where $D$ can be an order of magnitude larger than $D_0$, as well as it can lead to a paradox situation of a positive $V_0$, interpreted as an outward convection, despite a peaked flux surface averadged $n_z$ profile.

 Experimental values of $D$ and $V$ are often strongly correlated, thus challenging to separate.  This, together with a demand for a flexible low dimensional representation of transport coefficients profiles, motivated a search of their most efficient parameterization. The best results are obtained with profiles of  $\log(D)$ and $V\!/D$ ratio described by even and odd spline, respectively. Sharp and localized changes in transport coefficients in a narrow edge transport barrier (ETB) and close to the axis, together with simple profiles in between, calls for application of free-knot splines \cite{jupp1978approximation} where not just knot values but also knot locations are varied. Finally, since the close positioning of knots can lead to unphysical splines overshoots, we have utilized   Piecewise Cubic Hermite Interpolating Polynomial (PCHIP) splines \cite{fritsch1980monotone}, which preserves monotonicity in between knots and reaches extremes only in the locations of the knots. The pedestal pinch was parameterized by a Gaussian with amplitude $V_\mathrm{ped}$, width $w_\mathrm{ped}$, and position $\rho_\mathrm{ped}$ added to the $V\!/D$ spline to reduce the number of free parameters. 

 Transport outside of the separatrix and boundary condition are treated by a simplified 0D model in STRAHL\cite{dux2014strahl}. Assuming in addition to the radial transport also a fast parallel flux along open field lines and a finite level of recycling from a wall and divertor reservoir. This model depends on several parameters, namely the SOL Mach number, $\tau_\mathrm{div\rightarrow SOL}$,  $\tau_\mathrm{div\rightarrow pump}$, recycling coefficient $R$, SOL width, width of limiter shadow and kinetics and radial transport coefficients profile in SOL.  
 
 The last important transport mechanism included in the forward model are ELMs causing a sudden loss of pedestal in the impurity density.  Similarly to the previous studies\cite{putterich2011elm,dux2003chapter,janzer2015tungsten,casali2018modelling}, ELMs are modeled by a rapid increase of diffusion and reduced inward pinch in ETB for the ELM duration. Nevertheless, we have not modeled associated variations in background kinetics profiles.  
 Given a complex nature of ELMs and the essential role of parallel flow along field lines, this simplified model attempts to reproduce observed drop in the edge impurity density rather than the actual transport during the ELMs, which is out of the scope of this work.
 
 The forward model described in this section, as well as the uncertainties quantification presented in the next section, are publicly available as part of the open-source project OMFIT\cite{OMFIT2015} to help in the advancement of experimental impurity research in fusion devices.

\section{Bayesian inference of transport coefficients and quantification of uncertainties}

While the solution of the forward model is straightforward, the inversion task i.e., determining transport coefficients from experimental measurements, is an ill-posed non-linear problem. Ill-posed in this context means that the observations can be represented equally well by a wide range of transport coefficients.  
Furthermore,  the forward model depends on several nuance parameters of the edge model, which despite not being in a primary focus of this work,  play a significant role in impurity confinement, and the uncertainty of these parameters must be propagated to the uncertainty of the transport coefficients.

An approach to a solution of this inversion task, commonly applied in previous experimental studies \cite{giroud2007method,dux2003influence,valisa2011metal,sertoli2011local,howard2012quantitative,villegas2010experimental,odstrcil2017physics,zhang2016investigation}, is least-squares minimization of the difference between the forward model and observations. Under the assumption of independently distributed Gaussian noise in the observations, the uncertainty of the model parameters can be estimated from a diagonal of the covariance matrix \cite{press1992numerical}.  But unrealistically small uncertainties were reported \cite{carraro2007impurity,sertoli2011local,chilenski2018efficient}, likely as a consequence of systematical errors unaccounted in the measurements and too rigid model parameterization.  Several studies \cite{giroud2007method,dux2003influence,valisa2011metal} estimated uncertainty by varying a solution within a specific range of $\chi^2$, but this approach becomes particularly challenging as the number of model parameter increases or some are poorly constrained by available measurements.

A Bayesian approach, proposed in a statistical study \cite{chilenski2017experimental}, has a potential to overcome these issues.  A representation of the model parameters $\mathbf{\theta}$ is given by their joint \emph{posterior} distribution $f(\mathbf{\theta}|\mathbf{s})$ conditioned by an observation $\mathbf{s}$.    
A posterior distribution is related to the \emph{likelihood}  $f(\mathbf{s}|\mathbf{\theta})$ and \emph{prior}  $f(\mathbf{\theta})$   via Bayes' rule:
\begin{equation}
 f(\mathbf{\theta}|\mathbf{s}) = \frac{f(\mathbf{s}|\mathbf{\theta})  f(\mathbf{\theta})}{f(\mathbf{s})}.
\end{equation} 
 \emph{Likelihood}  $f(\mathbf{s}|\mathbf{\theta})$  express the probability of $\mathbf{s}$ to be observed given some values of the parameters $\mathbf{\theta}$, while \emph{prior} $f(\mathbf{\theta})$ represent any information available about $\mathbf{\theta}$ before observations $\mathbf{s}$ are included and $f(\mathbf{s})$ is \emph{evidence} which in this context acts as a normalization constant independent of $\theta$. The Bayesian inference of the transport coefficient is  performed by a marginalization over a nuisance parameter $\tilde{\theta}$ of the forward model
\begin{equation}
 f(\mathbf{D,V| s}) =  \int  f(\mathbf{D,V,\tilde{\theta}}|\mathbf{s}) \mathrm{d}\tilde{\theta}. 
\end{equation} 
This step is carried out using Markov chain Monte Carlo (MCMC) integration  by  a parallel-tempered affine-invariant ensemble sampler EMCEE\cite{goodman2010ensemble,earl2005parallel}. This algorithm runs multiple MCMC’s at different temperatures where each MCMC is an ensemble of many walkers performing a random walk guided by the posterior distribution to yield a set of random samples. Marginalization is then a trivial discarding of dimension corresponding to nuance parameters in these samples.

 Both prior and likelihood remain to be specified.  The prior distribution for knot values of transport coefficients was defined by a univariate distribution with boundaries well outside of the expected parameter range to minimize the introduced bias but sufficient to prevent numerical issues in STRAHL. Furthermore, the logarithmic transformation of the diffusion profile applied in the forward model is equivalent to a prior $ f(D) \propto 1/D $, which equalizes all scales, i.e. both small and large values of $D$ are equally likely. The knot positions are forced by a \emph{prior} to maintain a monotonous order and a minimal distance in between because too steep gradients in the transport coefficients can result in a numerical issue with particle conservation in STRAHL.  The actual prior distributions for each parameter are listed in Tab.~\ref{tab:prior}. 

\begin{table}[h!]
\centering
\caption{Table summarizing a prior definitions applied for Bayesian inference of transport coefficients parameterized by splines and other nuance parameters of the forward model}
\smallskip 
\label{tab:prior}
{\renewcommand{\arraystretch}{1.2}
\begin{ruledtabular}
\begin{tabular}{@{}ll@{}}
Parameter $\theta$ & prior $p(\theta)$\\\hline
$\log(D_i)$ &  $\mathcal{U}(-2,2)$\\
$(V\!/D)_i$ & $\mathcal{U}(-200,100)$ \\
$r_{D_i}$ & $\mathcal{U}(r_{D_{i-1}}+0.02,r_{D_{i+1}}-0.02)$ \\
$r_{V_i}$ & $\mathcal{U}(r_{V_{i-1}}+0.02,r_{V_{i+1}}-0.02)$ \\
$V_\mathrm{ped}$ & $\mathcal{U}(-200,0 )$ \\
$w_\mathrm{ped}$ & $\mathcal{U}(0.01,0.1 )$ \\
$\rho_\mathrm{ped}$ & $\mathcal{U}(0.95,1.05 )$ \\
$D_\mathrm{ELM}$ & $\mathcal{U}(D_n, 100)$ \\
$V_\mathrm{ELM,ped}$ & $\mathcal{U}(V_\mathrm{ped},100 )$ \\
$\tau_{div\rightarrow SOL}$ & $\mathcal{U}(1,100)$ \\
$\tau_{div\rightarrow pump}$ & $\mathcal{U}(1,1000)$ \\
$R$ & $\mathcal{U}(0,1)$ \\
$M_\mathrm{SOL}$ & $\mathcal{U}(0.01,0.2)$ \\

\end{tabular}
\end{ruledtabular}
}
\end{table} 

Assuming a Gaussian independently distributed noise in measurements $s_i$ with a variance $\sigma_i^2$ for each of $N$ available measurements and an accurate model $ {s_i}$, the log-likelihood is given up to an additive constant by
\begin{equation}
  -2 \ln(f(\mathbf{s}|\mathbf{\theta})) = \chi^2  = \sum_{i = 1}^N \frac{(s_i-\hat{s}_i(\theta))^2}{\sigma_i^2}. 
\end{equation} 
However, the uncertainty of the transport coefficients inferred using this likelihood are in an order of just a few percent, significantly less than expected from basic sensitivity studies. A possible explanation is an underestimated variance
  $\sigma_i$. Therefore, $\chi^2_k$ of each diagnostic was  rescaled by a factor $N_k/\chi^2_{\mathrm{MAP},k}$ calculated from  maximum a posteriori
 (MAP) estimate. Nevertheless,  the factors $\chi^2_{\mathrm{MAP},k}/N_k$ are of the order of one, insufficient to significantly change the uncertainty. 

Questionable is also the assumption of independently distributed errors.  As the main reasons for the correlation are systematic uncertainties in calibrations, geometry, plasma position,  atomic data, simplified forward model, small background variation as well as low frequency noise in the measured signals. Reduction in information content of the correlated signal is quantified by a effective sample size (ESS)  \cite{ripley2009stochastic}, equivalent to number of samples $N_\mathrm{eff}$ from uncorrelated signal required to achieve the same level of precision. Assuming that residuum $\mathbf{r}_\mathrm{MAP} \equiv \mathbf{s}-\hat{\mathbf{s}}$ of MAP estimate is represented by the  autoregressive process of the first order with a parameter $\lambda$ and an autocorrelation time $\tau_\mathrm{ac} = -1/\ln \lambda$; $N_\mathrm{eff}$ is given by a formula 
\begin{equation}
N_\mathrm{eff} = 1+(N-1) \frac{1-\lambda}{1+\lambda}.  
\label{eq:lambda}
\end{equation} 
where for $\tau_\mathrm{ac} \gg 1$,  $N_\mathrm{eff} \approx N/(2\tau_\mathrm{ac})$. The parameter $\lambda$ is estimated from $N_z$ -- number of times $\mathbf{r}_\mathrm{MAP}$  up-crosses level zero\cite{johannesson2016ar}
\begin{equation}
 \lambda = \cos(2 \pi N_z/N). 
\end{equation} 
Independently distributed noise has is $N_z\approx N/4$, therefore $\lambda\approx 0$ and following Eq.~\eqref{eq:lambda} also $N_\mathrm{eff}\approx N$. Because ESS is different for every diagnostic $k$, each $\chi^2_k$ was scaled by $N_{\mathrm{eff},k}/N_k$ independently, which is virtually equivalent to downsampling measured signals by a factor  $N_k/N_{\mathrm{eff},k}$. Diagnostics scaling factors 
\begin{equation}
\alpha_k = \frac{N_\mathrm{eff,k}}{N_k}\frac{N_k}{\chi^2_{\mathrm{MAP},k}}
\end{equation}
 are properly weight diagnostics with respect each other and increase uncertainty by an order of magnitude. Since the MAP solution also depends on $\alpha_k$, scaling factors for likelihood are be determined iteratively before the actual Bayesian inference is executed.

One of the reasons why the Bayesian inference via MCMC was not applied in previous experimental impurity transport studies is the enormous computational cost.
A single STRAHL evaluation needs 1\,s of computation time, and thus for 400 chains, 5 temperatures, and $\sim 30000$ iteration to converge, generating samples from posterior distribution requires about 7 wall-clock days on 32 cores cluster.
To accelerate the convergence, the actual chains are initialized from a small sphere around the final MAP estimate because initialization from a prior distribution is not a computationally tractable approach.
 Furthermore, given the computational cost of the  Bayesian inference, the optimal number of spline knots was not determined based on a maximum of \emph{evidence} as proposed in Ref.~\onlinecite{chilenski2017experimental}, but instead it was found from a MAP estimate as the lowest needed to reproduce experimental data plus one. The additional knot provides MCMC more freedom to explore the parameter space of possible solutions.

\bibliographystyle{revtex}

\bibliography{paper}
\end{document}